\documentclass[manuscript]{aastex61}

\usepackage{color}
\usepackage{mathrsfs}
\usepackage{amsmath}
\usepackage{lineno}
\usepackage{graphicx}
\usepackage{float}

\def\VEC#1{\mbox{\boldmath $#1$}}

\shorttitle{1D linear analysis and simulation of Alfv\'{e}n wave around a Kerr black hole}
\shortauthors{S. Koide, S. Noda, \& M. Takahashi}

\begin{document}


\title{One-dimensional linear analysis and numerical simulations of Alfv\'{e}n waves
in a force-free magnetosphere around a Kerr black hole}

\vspace{1cm}


\author{Shinji Koide}
\affil{Department of Physics, Kumamoto University, 
2-39-1, Kurokami, Kumamoto, 860-8555, JAPAN}
\email{koidesin@kumamoto-u.ac.jp}
\author{Sousuke Noda}
\affil{National Institute of Technology, Miyakonojo College, Miyakonojo 885-8567, Japan}
\author{Masaaki Takahashi}
\affil{Department of Physics and Astronomy, Aichi University of Education,
Kariya, Aichi 448-8542, Japan}



\begin{abstract}
We perform one-dimensional linear analysis and numerical simulations of 
the propagation of Alfv\'{e}n waves in a force-free magnetosphere
along magnetic field lines around a spinning black hole.
We use the results to investigate the dynamic process of wave propagation and 
energy transport for Alfv\'{e}n waves.
As in a previous study using the Banados--Teitelboim--Zanelli 
spacetime \citep{koide22},
the Alfv\'{e}n wave induces a fast magnetosonic wave in the case of a spinning black hole.
Energy conservation is confirmed when this additional induced magnetosonic wave is considered.
We also observe 
the reflection of the inwardly propagating Alfv\'{e}n wave
around the static limit, which is prohibited in theory when using the eikonal 
approximation.
\end{abstract}



\section{Introduction} \label{sec1}

The centers of galaxies classified as active galactic nuclei (AGNs)
are accompanied by drastic phenomena like the superluminal motion of knots in
radio observations and flares in near-infrared, X-ray, and $\gamma$-ray 
observations \citep{pearson81, biretta99, acciari09, gravity17}.
The superluminal motion of knots is explained by inhomogeneous emission
from relativistic jets directed close to us and the flares
are expected to be rapid energy releases from
the magnetic field due to
magnetic reconnection.
These drastic phenomena are explained by interactions between
plasma and magnetic fields around the supermassive black holes
enshrined at the centers of AGNs.

The simplest approximation of the interaction between the plasma and magnetic fields
is given by magnetohydrodynamics (MHD).
General relativistic MHD (GRMHD) simulations have shown that 
the interaction between the plasma and magnetic field around black holes
causes complex phenomena, including the formation of relativistic jets
and flares 
\citep{koide06,mckinney06,porth19,ripperda20,nathanail20,hakobyan23}.
These complex phenomena are made up of many elementary processes 
of plasma physics and the general theory
of relativity. For example, GRMHD simulations have shown that the relativistic jets
are powered by the Blandford--Znajek mechanism \citep{blandford77}, where the jet
driving energy comes from the extracted rotational energy of the spinning black hole
as a torsional Alfv\'{e}n wave.

A similar phenomenon of energy extraction from spinning black holes 
is caused by wave scattering.
That is, a wave-like scalar field, electromagnetic waves, and
gravitational waves extract the rotational energy of the black hole
in a process called ``superradiance"\citep{press72,teukolsky74}. 
MHD waves also cause superradiance.
In fact, within the force-free approximation, 
the magnetosonic fast wave, one of MHD waves, behaves like a scalar wave
at light speed, and
the fast magnetosonic wave can cause superradiance \citep{uchida97b}.
The other mode of MHD waves, the Alfv\'{e}n wave, may also cause superradiance
in the force-free limit.
\citet{uchida97a,uchida97b} investigated the
Alfv\'{e}n wave around a black hole
under the force-free condition.
He concluded that the superradiance of the
Alfv\'{e}n wave never happens within the eikonal limit because
there is no reflection of the Alfv\'{e}n wave near the ergosphere.
\citet{noda20,noda22} derived a linear equation for a
force-free Alfv\'{e}n wave
propagating along a magnetic field line around 
a Banados--Teitelboim--Zanelli (BTZ) black string and a Kerr black hole.
They discussed the wave scattering problem as seen by an observer
who rotates with a magnetic field line (a corotating observer).
By solving a stationary wave scattering problem with a finite (or long) wavelength,
they concluded that the reflection rate observed by the corotating observer can
exceeded unity, representing superradiance, under Blandford-Znajek process conditions.
To determine the reflection rate, they used the asymptotic solution of the linear
equation of the Alfv\'{e}n wave with Wronskian conservation.

To explore the details of superradiance energy transport in an
Alfv\'{e}n wave, a distinct solution around the reflection region is required.
This motivates us to conduct additional linear analysis and 
a numerical study of an Alfv\'{e}n wave around a black hole.
As a first step, we study the wave propagation around a BTZ black string
numerically \citep{koide22}.
The numerical calculations reveal some interesting results regarding
the dynamic processes of Alfv\'{e}n waves.
However, the BTZ spacetime corresponds to an infinitesimally long black string 
with a singularity at an infinite distance from the string, which is 
incompatible with real astrophysical objects.

\begin{figure} 
\begin{center}
\includegraphics[scale=0.8]{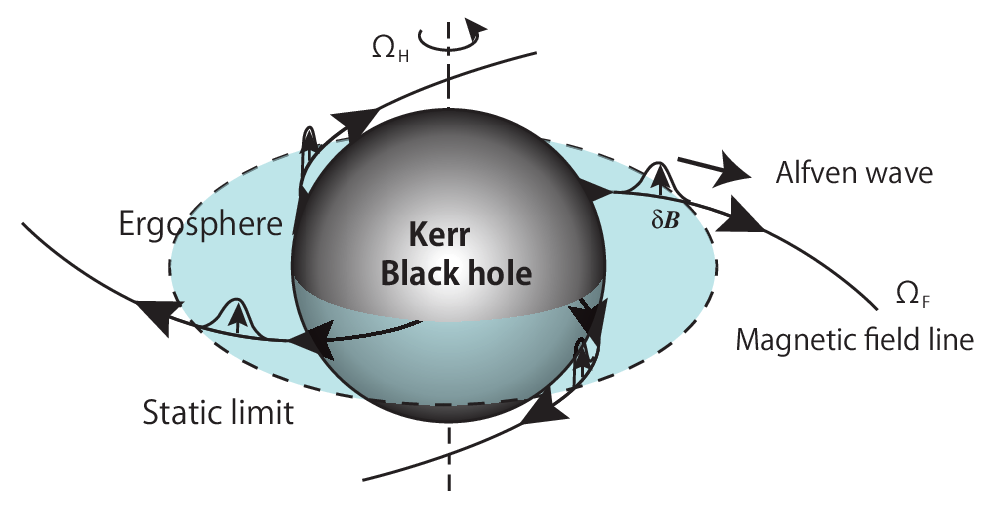}
\end{center}
\caption{A schematic of axisymmetric Alfv\'{e}n wave pulse propagations along 
magnetic field lines on the equatorial plane
around a Kerr black hole.
The dashed ring shows the static limit, which is the boundary of the ergosphere.
$\Omega_{\rm F}$ and $\Omega_{\rm H}$ are constants along the 
stationary magnetic field line.
$\Omega_{\rm F}$ corresponds to the angular velocity of the stationary magnetic field line,
and $\Omega_{\rm H}$ is the angular velocity of the normal frame at the horizon.
\label{pontie_kerralfven}}
\end{figure}

In this paper, we investigate
Alfv\'{e}n wave propagation
in the equatorial plane of a Kerr black hole (Fig. \ref{pontie_kerralfven}). 
With proper coordinates, which include a spatial coordinate along the magnetic
field line, the wave equation for the Alfv\'{e}n wave becomes simple and similar to the 
wave equation of a string.
In some parameter ranges, the string corresponding to the equation
is slightly strange, for example, with negative tension
outside the region between the two light surfaces. 
This strange equation reflects surprising and interesting dynamics of
the Alfv\'{e}n wave around a Kerr black hole.
For example, the Alfv\'{e}n wave is unstable in the negative tension region,
and even in the stable case, the energy of the Alfv\'{e}n wave is not conserved.
It is shown in this work that including the contribution of a fast magnetosonic wave
induced by the
Alfv\'{e}n wave conserves energy as in the case of the BTZ metric
\citep{koide22}.
The induction of a fast magnetosonic wave by an Alfv\'{e}n wave was 
also investigated
by \citet{yuan21} for a different magnetic field configuration around a pulsar
(neutron star) using special relativistic force-free electromagnetic dynamics
to explain the possible energy loss mechanism in the pulsar magnetosphere.
The authors suggested that the timescale of dissipation of the Alfv\'{e}n wave to a fast magnetosonic wave
in the closed magnetosphere around a neutron star should be comparable 
to the duration of the X-ray bursts of some pulsars.

This paper is organized as follows.
In Section \ref{secalf}, we explain the linear theory for
Alfv\'{e}n waves using first-order perturbation theory
along stationary background magnetic field lines around a black hole.
The second-order variable $\chi$ is introduced, which describes
the fast magnetosonic wave induced by the Alfv\'{e}n wave.
In Section \ref{sec3}, we investigate the conservation of energy and angular momentum 
for the Alfv\'{e}n wave and the induced fast magnetosonic wave.
In Section \ref{secmet}, we summarize the numerical methods, and
in Section \ref{secres}, we present the numerical results.
We comprehensively summarize and discuss the results in Section \ref{secdis}.
Throughout this paper, 
we use the natural unit system, where the speed of light and gravitational constant
are unity (i.e., $c=1$ and $G=1$).

\newpage

\section{Background magnetosphere and Alfv\'{e}n wave
\label{secalf}}

We investigate the propagation of the Alfv\'{e}n wave along a magnetic field line
in a force-free magnetosphere
around a Kerr black hole.  
Under the force-free condition $J^\mu F_{\mu\nu}=0$ 
($J^\mu$ is the 4-current density and $F_{\mu\nu}$ is the electromagnetic field tensor),
with the homogeneous Maxwell equation $\nabla_\mu {}^* F^{\mu\nu}=0$
(${}^* F^{\mu\nu}$ is the dual of $F_{\mu\nu}$; ${}^* F^{\mu \nu} = \epsilon^{\mu\nu\rho\sigma} F_{\rho\sigma}/2$),
the electromagnetic field tensor $F_{\mu\nu}$ 
is represented
by the Euler potentials $\phi_1$ and $\phi_2$ as:
\begin{equation}
F_{\mu \nu}  = \partial_\mu \phi_1 \partial_\nu \phi_2 
- \partial_\nu \phi_1 \partial_\mu \phi_2 
\label{defphi1phi2}
\end{equation}
as shown by
\citet{uchida97a,uchida97b} \textcolor{black}{and used by \citet{noda20,noda22}}.
\textcolor{black}{The contour surfaces of the Euler potentials $\phi_1$ and $\phi_2$ 
at a certain time
give the magnetic surfaces at the time as shown below.}
Here, $\epsilon^{\mu\nu\rho\sigma}$ is the completely antisymmetric tensor
and determined as $\epsilon^{0123}=1/\sqrt{-g}$
and $g$ is the determinant of the metric tensor $g_{\mu\nu}$, with
$g \equiv \det(g_{\mu\nu})$.
\footnote{The sign of $\epsilon^{\mu\nu\rho\sigma}$ is opposite
to that of \citet{noda22}}
With Euler potentials, the homogeneous Maxwell equation 
$\nabla_\mu {}^* F^{\mu\nu}=0$ 
is satisfied identically. 
Using the inhomogeneous Maxwell equation $\nabla_\lambda F^{\lambda \mu} = - J^\mu$ 
and the condition of the force-free field $J^\mu F_{\mu\nu} =0$ with 
Eq. (\ref{defphi1phi2}),
we obtain equations for $\phi_1$ and $\phi_2$:
\begin{equation}
\partial_\lambda \phi_i \partial_\nu [
\sqrt{-g} W^{\lambda\alpha\nu\beta}
\partial_\alpha \phi_1 \partial_\beta \phi_2 ] = 0  \verb!   !(i=1,2),
\label{nodaeq04w}
\end{equation}
where $W^{\lambda\alpha\nu\beta} = g^{\lambda\alpha} g^{\nu\beta}
- g^{\lambda\beta} g^{\alpha\nu}$.
\footnote{We note the following useful formulae concerning $W^{\lambda\alpha\nu\beta}$:
$ W^{\lambda\alpha\mu\beta} = W^{\alpha\lambda\beta\mu} = W^{\beta\mu\alpha\lambda} 
= W^{\mu\beta\lambda\alpha} 
=-W^{\lambda\beta\mu\alpha} = -W^{\mu\alpha\lambda\beta}$ ,
$ W^{\mu\nu\lambda\kappa} + W^{\mu\kappa\nu\lambda} + W^{\mu\lambda\kappa\nu} = 0$.
}
For intuitive discussion in the following sections, we use the electric field
$\VEC{E}$ and the magnetic field $\VEC{B}$.
The 4-electric field and 4-magnetic field are defined as
$ E^\lambda = F^{\lambda \nu} N_\nu =  - \alpha F^{\lambda 0} 
= - \alpha (\partial^\lambda \phi_1 \partial^0 \phi_2 
- \partial^0 \phi_1 \partial^\lambda \phi_2 )$  and
$B^\lambda = {}^* F^{\lambda \nu} N_\nu =  - \alpha {}^* F^{\lambda 0} = 
\frac{1}{\sqrt{\gamma}} \eta^{0 \lambda \alpha \beta}
\partial_\alpha \phi_1 \partial_\beta \phi_2 $, respectively,
where $\eta^{\alpha\beta\gamma\delta}$ is the Levi-Civita symbol, with $\eta^{0123}=1$
and
$N^\mu$ is the 4-velocity of the normal observer frame $x^{\tilde{\mu}}$.
The spatial components of $E^\mu$ and $B^\mu$ give the components of the 
electric field $\VEC{E}$ and the magnetic field $\VEC{B}$.
\textcolor{black}{
The contour surface $\phi_m$ ($m=1, 2$) at a certain time gives the magnetic surface
at the time
because the surface is tangent to $\VEC{B}$: $B^i \partial_i \phi_m 
= B^\lambda \partial_\lambda \phi_m 
= \frac{1}{\sqrt{\eta}} \eta^{0\lambda\alpha\beta} \partial_\alpha \phi_1
\partial_\beta \phi_2 \partial_\lambda \phi_m  = 0$.}

\subsection{The force-free stationary background solution in Boyer--Lindquist coordinates}
The metric of the Boyer--Lindquist coordinates $x^\mu_{\rm BL} = (t, r,\theta, \varphi)$
is given by 
\begin{equation}
ds^2 = g_{\mu\nu}^{\rm BL} dx^\mu_{\rm BL} dx^\nu_{\rm BL} 
= - \alpha^2 dt^2 + \gamma_{ij} (dx^i_{\rm BL} + \beta^i dt) (dx^j_{\rm BL} + \beta^j dt) ,
\end{equation}
where
Greek indices such as $\mu$ and $\nu$ run from 0 to 3,
while Roman indices such as $i$ and $j$ run from 1 to 3,
$\displaystyle \alpha^2 = - g_{tt} + \gamma_{ij} \beta^i \beta^j 
= \frac{\Sigma \Delta}{A}$,
$\displaystyle \gamma_{rr}=\frac{\Sigma}{\Delta}$,
$\displaystyle \gamma_{\theta\theta}=\Sigma$,
$\displaystyle \gamma_{\phi\phi}=\frac{A}{\Sigma} \sin^2 \theta \equiv R^2$,
$\displaystyle \gamma_{ij}=0, (i \neq j)$,
$\displaystyle \beta^\phi=-\Omega$,
$\displaystyle \beta^r = \beta^\theta=0$,
$\displaystyle \Delta = r^2 - 2 M r + a^2$, 
$\displaystyle \Sigma = r^2 + a^2 \cos^2 \theta$,
$\displaystyle A = (r^2 + a^2)^2 - a^2 \Delta \sin^2 \theta$,
and
$\displaystyle \Omega = - \frac{g_{r \varphi}^{\rm BL}}{g_{\varphi\varphi}^{\rm BL}} = \frac{2 M a r}{A}$. 
Here, the spin parameter of a black hole $a$ is given by $a = J/M$, where
$M$ and $J$ are the mass and angular momentum of the black hole, respectively.
The \textcolor{black}{dimensionless} spin parameter $a_\ast$ is given by $a_\ast=a/M$.
The determinant of the metric tensor for 
the Boyer--Lindquist coordinates $(g_{\mu\nu}^{\rm BL})$ is given by 
$g_{\rm BL} \equiv \det(g^{\rm BL}_{\mu\nu}) = - \Sigma^2 \sin^2 \theta$, and
the 4-velocity of the normal observer 
\textcolor{black}{(Zero-Angular-Momentum Observer: ZAMO)} is given by
\begin{equation}
N^\lambda = \left (\frac{1}{\alpha}, - \frac{\beta^i}{\alpha} \right ),
N_\lambda = (- \alpha, 0, 0, 0).
\end{equation}
Using Eq. (\ref{nodaeq04w}), we obtain
the steady-state solution of the force-free magnetic field around the equatorial plane 
($\theta \sim \pi/2$) of the black hole
\begin{equation}
\bar{\phi}_1 =  - \cos \theta, \verb!   !
\bar{\phi}_2 = 
\int \frac{I r^2}{\Delta} dr + \varphi - \Omega_{\rm F} t,
\label{eqequisol}
\end{equation}
where $\bar{\phi}_i$ ($i=1,2$) represents the steady-state solution,
and $\Omega_{\rm F}$ and $I$ are constants along a field line. 
These constants mean that
$\Omega_{\rm F}$ is the angular velocity of the magnetic field line, and
$I$ is the total current, which is obtained
from the regularity of $F^{\mu\nu} F_{\mu\nu}$ at the black hole horizon as,
\begin{equation}
I = \frac{2 M}{r_{\rm H}} (\Omega_{\rm H} -\Omega_{\rm F}).
\label{cond4i}
\end{equation}
The solution given by Eq. (\ref{eqequisol}) corresponds to the field of the magnetic monopole
around the equatorial plane.
Here, the strength of the stationary magnetic field is normalized without loss of generality.
\textcolor{black}{
Note that $\phi_m$ ($m=1, 2$) corresponds to the stream function and the contour surface
of $\bar{\phi}_m$ at a certain time represents the steady-state magnetic surfaces.
In the present case, $\bar{\phi}_1 = - \cos \theta= 0$ represents 
the steady magnetic surface along the equatorial plane.
The magnetic flux surface $\bar{\phi}_2 =$ const. is vertical to the equatorial plane
and contains the magnetic field line on the equatorial plane.}\\
%
The corotating Killing vector of the field line, 
$\xi_{(\rm F)}^\nu \equiv \xi_{(t)}^\nu + \Omega_{\rm F} \xi_{(\varphi)}^\nu$ is
parallel to the 4-velocity of an observer corotating with  the magnetic field line,
where $\xi^\nu_{(t)}$ is the Killing vector in the time direction and $\xi^\nu_{(\phi)}$ 
is the Killing vector in the azimuthal direction.
The norm of $\xi_{(\rm F)}^\nu$ is
\begin{equation}
\Gamma \equiv g_{\mu\nu}^{\rm BL} \xi^\mu_{\rm (F)} \xi^\nu_{\rm (F)}
=- \alpha^2 + R^2 (\Omega - \Omega_{\rm F})^2 ,
\end{equation}
where the roots of $\Gamma=0$ give the location of the inner and outer light surfaces.

\subsection{Magnetic natural frame}

To simplify the form of the calculations, 
we utilize the coordinates $(T, X, \rho, z)$, defined as
\begin{align}
T &= t , & \\ 
X &= r , &  \\ 
\rho &= \varphi - \int \frac{I X^2}{\Delta}dr , & \\ 
  \Psi &= - \cos \theta . &   
\end{align}
The coordinate frame does not rotate, although it is similar to the coordinate frame 
used by \citet{noda22}, which rotates with the same angular velocity as the
magnetic field lines.  
In this coordinate system,
$X$ is the coordinate along the magnetic field line, so we call this
coordinate system $(T, X, \Psi, \rho)$ the ``magnetic natural frame".
The contravariant component of the metric tensor of the magnetic natural frame 
$(T, X, \Psi, \rho)$ is as follows:
\begin{equation}
(g^{\mu\nu}) = \left ( \begin{array}{cccc}
g^{tt}_{\rm BL} & 0 & 0 & g^{\varphi t}_{\rm BL} \\
0 & g^{rr}_{\rm BL} & 0 & \displaystyle \frac{I X^2}{\Delta} g^{rr}_{\rm BL} \\
0 & 0 & g^{\theta\theta}_{\rm BL} \sin^2 \theta & 0 \\
g^{\varphi t}_{\rm BL} & \displaystyle \frac{I X^2}{\Delta} g^{rr}_{\rm BL} & 0 &\displaystyle g^{\varphi\varphi}_{\rm BL} + \left (\frac{I X^2}{\Delta} \right )^2 g^{rr}_{\rm BL}\end{array} \right )
= \left ( \begin{array}{cccc}
\displaystyle - \frac{1}{\alpha^2} & 0 & 0 & \displaystyle - \frac{\Omega}{\alpha^2} \\
0 & \displaystyle \frac{\Delta}{\Sigma} & 0 & \displaystyle \frac{I X^2}{\Sigma} \\
0 & 0 & \displaystyle \frac{\sin^2 \theta}{\Sigma} & 0 \\
\displaystyle - \frac{\Omega}{\alpha^2} & \displaystyle \frac{I X^2}{\Sigma} & 0 & \displaystyle \frac{1}{R^2} - \frac{\Omega}{\alpha^2} + \frac{I^2 X^4}{\Delta \Sigma} \end{array} \right ).
\label{gupmunu}
\end{equation}
The determinant of the matrix $(g_{\mu\nu})$ is given by $g = \det (g_{\mu\nu}) = - \Sigma^2$.
In the magnetic natural frame, the steady-state force-free solution
given by Eq. (\ref{eqequisol}) is simply
\begin{equation}
\bar{\phi}_1 = \Psi, \verb!   !
\bar{\phi}_2 = \rho - \Omega_{\rm F} T.
\label{eqequsolconaf}
\end{equation}


\subsection{Wave equation for the Alfv\'{e}n wave}



In the background magnetosphere given by Eq. (\ref{eqequsolconaf}), 
we consider a perturbation of the magnetic field; that is, 
the propagation of Alfv\'{e}n waves.
We already know that an Alfv\'{e}n wave induces a fast wave represented by a second-order
perturbation, as suggested by the calculations in BTZ spacetime 
\citep{koide22}.
Therefore, we should consider not only the Alfv\'{e}n wave but also the fast wave using
second-order perturbation.
The infinitesimally small perturbation to the Euler potential 
$\phi_i \longrightarrow \bar{\phi}_i + \delta \phi_i$
is given by the displacement (polarization) vector $\zeta^\lambda$ as
$\delta \phi_i = \zeta^\lambda \partial_\lambda \phi_i$.
First, we consider the Alfv\'{e}n wave with a first-order perturbation.
We focus on a pure Alfv\'{e}n wave propagating on the magnetic surface 
at the equatorial plane ($\Psi = - \cos \theta =0$)
and oscillating perpendicularly to the magnetic surface:
$\delta \zeta^\lambda =(0, 0, \delta \zeta^\Psi, 0)$. 
\textcolor{black}{This wave is the ``pure" Alfv\'{e}n wave because the polarization direction
of the oscillation $\delta \zeta^\lambda$ is perpendicular to that of the propagation 
which is parallel to the equatorial plane $\Psi = \bar{\phi}_1 = 0$.}
Then, the perturbed Euler potentials are
\begin{equation}
\phi_1 = \bar{\phi}_1 + \delta \phi_1 = \Psi + \delta \phi_1, \verb!   ! 
\phi_2 = \bar{\phi}_2 = \rho - \Omega_{\rm F} T,
\end{equation}
because $\delta \phi_2 = \delta \zeta^\mu \partial_\mu \bar{\phi}_2 
= \delta \zeta^\Psi \partial_\Psi (\rho - \Omega_{\rm F} T) = 0$.
Hereafter, we denote the stationary background (zeroth-order) term, first-order perturbation, 
second-order perturbation, $\cdots$ of a variable $A$ as
$\bar{A}$, $\delta A$, $\delta^2 A$, $\cdots$, i.e., $A = \bar{A} + \delta A + \delta^2 A
+ \cdots$.
Without loss of generality, we assume the perturbation for the Alfv\'{e}n wave 
is axisymmetric, i.e., the perturbation is independent of $\rho$:
\begin{eqnarray}
\delta \phi_1 &=& \psi (T, X, \Psi) ,
\end{eqnarray}
where the Alfv\'{e}n wave propagates along a magnetic field line
and never interacts with Alfv\'{e}n waves propagating along other magnetic field lines.

In nonrotating natural coordinates, the linearization of Eq. (\ref{nodaeq04w}) with $i=2$ yields
\begin{equation}
\partial_\lambda (\sqrt{-g} W^{\lambda\alpha\mu\beta}
\partial_\alpha \psi \partial_\beta \bar{\phi}_2 ) \partial_\mu \bar{\phi}_1
+ \partial_\lambda (\sqrt{-g} W^{\lambda\alpha\mu\beta} 
\partial_\alpha \bar{\phi}_1 \partial_\beta \bar{\phi}_2 ) \partial_\mu \psi = 0,
\label{eq4i2} 
\end{equation}
and can be calculated easily as
\begin{equation}
\partial_\lambda (\sqrt{-g} Z^{\lambda\alpha} \partial_\alpha \psi) = 0,
\label{nodaeq16d}
\end{equation}
where $Z^{\lambda\alpha}
= W^{\lambda\alpha\mu\beta} \partial_\mu \bar{\phi}_2 \partial_\beta \bar{\phi}_2
= W^{\lambda\alpha\rho\rho}-\Omega_{\rm F}
(W^{\lambda\alpha\rho T}+W^{\lambda\alpha T\rho}) + \Omega_{\rm F}^2 W^{\lambda\alpha TT}$.
\footnote{$Z^{\mu\nu}$ is proportional to the projection operator onto 
the constant-$\bar{\phi}_2$ hypersurface, 
${\cal P}^{\mu\nu} \equiv g^{\mu\nu} - \partial^\mu \bar{\phi}_2
\partial^\nu \bar{\phi}_2/|\partial \bar{\phi}_2|^2 = 
|\partial \bar{\phi}_2 |^{-2} Z^{\mu\nu}$,
where $|\partial \bar{\phi}_2 | = \sqrt{\partial_\mu \bar{\phi}_2 \partial^\mu 
\bar{\phi}_2}$.}
Using the variables
$\lambda = - \sqrt{-g} Z^{TT}$, $S= \sqrt{-g} Z^{XX}$, $K= \sqrt{-g} Z^{\Psi\Psi}$,
and $V= -\sqrt{-g} Z^{TX}$, we rewrite Eq. (\ref{nodaeq16d}) as
\begin{equation}
\lambda \frac{\partial^2 \psi}{\partial T^2}
- \frac{\partial}{\partial X} \left ( S \frac{\partial \psi}{\partial X} \right ) 
+ \frac{\partial}{\partial T} \left ( V \frac{\partial \psi}{\partial X} \right ) 
+ \frac{\partial}{\partial X} \left ( V \frac{\partial \psi}{\partial T} \right ) 
+ K \psi = 0.
\label{newtontoyeq}
\end{equation}
This equation is identified by the equation of displacement of a string 
with a nonuniform line density $\lambda$ and nonuniform tension 
$\displaystyle S + \frac{V^2}{\lambda}$,
which moves with a nonuniform velocity $\displaystyle \frac{V}{\lambda}$ and is bounded 
by springs
with a spring constant per unit length $K$.
\footnote{Because Eq. (\ref{newtontoyeq}) is expressed by
$\displaystyle \lambda \left ( \frac{\partial}{\partial T} + \frac{V}{\lambda} 
\frac{\partial}{\partial X} \right )^2 \psi -  \frac{\partial}{\partial X}
\left [ \left ( S + \frac{V^2}{\lambda} \right ) \frac{\partial \psi}{\partial X} 
\right ] + K \psi + \frac{\partial V}{\partial X} 
\left ( \frac{\partial \psi }{\partial T} + \frac{V}{\lambda} \frac{\partial \psi}{\partial X} \right )
= 0$. When we neglect the nonuniformity of $V$, we obtain
the equation of a moving string,
$\displaystyle \lambda \left ( \frac{\partial}{\partial T} + \frac{V}{\lambda} 
\frac{\partial}{\partial X} \right )^2 \psi -  \frac{\partial}{\partial X}
\left [ \left ( S + \frac{V^2}{\lambda} \right ) \frac{\partial \psi}{\partial X} 
\right ] + K \psi = 0$.}
Using the metric tensor of the nonrotating natural coordinates (Eq. (\ref{gupmunu})) 
and the nature
of the perturbation $(\partial_\Psi^2 \psi = - \psi)$, we obtain the explicit expressions
for the parameters of the analogous string,
\begin{eqnarray}
\lambda &=& \frac{1}{\alpha^2} \left (\frac{\Sigma}{R^2}+\frac{I^2 X^4}{\Delta} \right ), \\
S &=& \frac{1}{\sin^2 \theta} [\alpha^2 - R^2(\Omega_{\rm F} - \Omega)^2] 
= \frac{-\Gamma}{\sin^2 \theta} ,
\label{exp2t} \\
K &=& \frac{1}{\Delta} \left [\alpha^2 - R^2(\Omega_{\rm F} - \Omega)^2 
+ \frac{I^2 X^4 \sin^2 \theta}{\Sigma} \right ] 
= \frac{\sin^2 \theta}{\Delta} \left ( S + \frac{I^2 X^2}{\Sigma} \right ) ,\\
V &=& \frac{I X^2}{\alpha^2} (\Omega_{\rm F} - \Omega),
\end{eqnarray}
where we have the identity $\displaystyle V^2 + \lambda S =\frac{\Sigma K}{\sin^2 \theta}$.
Equation (\ref{exp2t}) shows that the string analogous to Eq. (\ref{newtontoyeq}) 
is strange because the tension 
$\displaystyle S+\frac{V^2}{\lambda} = \frac{\Sigma}{\lambda} K$ 
becomes negative outside the
region between the inner and outer light surfaces, as $S = -\Gamma$ at the
equatorial plane.


\subsection{Dispersion relation of the Alfv\'{e}n wave and instability
\label{disrelalfwvinst}}

In the eikonal limit, that is, the short wavelength limit, 
Eq. (\ref{newtontoyeq}) yields 
the dispersion relation of the Alfv\'{e}n wave, Eq. (\ref{disrel4alfwv}).
Equation (\ref{disrel4alfwv})
reveals not only the propagation of the Alfv\'{e}n wave but also the instability
of the wave outside the outer light surface.
In Appendix \ref{appdx_disprel}, it is shown that the Alfv\'{e}n wave
is stable over the entire radial range ($r_{\rm H} < X < \infty$) when
$\displaystyle \Omega_{\rm F} < \frac{2M}{2M + r_{\rm H}} |\Omega_{\rm H}| \equiv \Omega_{\rm c}$.
Otherwise ($\Omega_{\rm F} > \Omega_{\rm c}$), the Alfv\'{e}n wave is unstable
at $X > r_{\rm m}$, where $X = r_{\rm m}$ is the solution for $K=0$.
In the numerical calculations in this paper, we use the parameters of the case where the Alfv\'{e}n
wave is stable over the entire radial range ($\Omega_{\rm F} < \Omega_{\rm c}$).

\subsection{Perturbations of the Alfv\'{e}n wave and induced wave
\label{sec_peralfwindw}}

\begin{figure}[H]
\begin{center}
\includegraphics[scale=0.8]{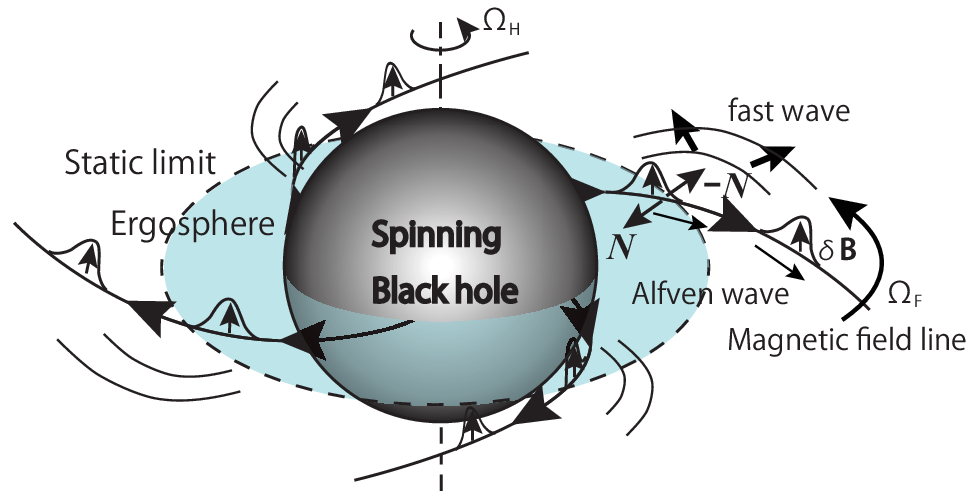}
\end{center}
\vspace{1cm}
\caption{A schematic of the Alfv\'{e}n wave and the wave induced by the
Alfv\'{e}n wave propagation along 
curved rotating field lines around a spinning black hole.
$N$ shows the force exerted by the background magnetic field line
on the Alfv\'{e}n wave. The reaction of the force produces an induced wave
oscillating in the azimuthal direction of the magnetic field line.
\label{pontie_indwav}}
\end{figure}

The induction of a fast wave by an Alfv\'{e}n wave is explained by a relativistic mechanism,
where the angular momentum of the Alfv\'{e}n wave 
changes in time (Fig. \ref{pontie_indwav}).
The Poynting flux of the Alfv\'{e}n wave is directed in the propagation direction
of the wave. The Poynting flux is proportional to the electromagnetic momentum
density of the Alfv\'{e}n wave.
Then, generally speaking, the Alfv\'{e}n wave propagating along the magnetic field line 
has its angular momentum as a second-order perturbation. 
Therefore, if the magnetic field line is curved azimuthally ($I \ne 0$)
or rotates ($\Omega_{\rm F} \ne 0$),
a torque on the
Alfv\'{e}n wave is required so that the Alfv\'{e}n wave traces the magnetic field line.
This means the force-free condition is broken at second order 
and the azimuthal component of the Lorentz force becomes finite.
To change the angular momentum
of the Alfv\'{e}n wave, the Alfv\'{e}n wave should be subject to an external 
torque to trace the rotating magnetic field line.
Thus, we take into account
additional first- and second-order perturbations $\delta^2 \phi_1$, 
$\delta^2 \phi_2$ as
\begin{eqnarray}
\phi_1 &=& \bar{\phi}_1 + \delta \phi_1 = \Psi + \psi(T, X, \Psi), 
\label{phi1per2nd} \\
\phi_2 &=& \bar{\phi}_2 + \delta^2 \phi_2 = \rho - \Omega_{\rm F} T + \chi(T, X, \Psi),
\label{phi2per2nd}
\end{eqnarray}
where $\chi \equiv \delta^2 \phi_2 = \delta^2 \zeta^\lambda \partial_\lambda \bar{\phi}_2
= \delta^2 \zeta^\lambda \partial_\lambda (\rho - \Omega_{\rm F} T)
= \delta^2 \zeta^\rho - \Omega_{\rm F} \delta^2 \zeta^T$.
\textcolor{black}{
This perturbation corresponds to the fast wave because the polarization of the 
oscillation of the magnetic surface $\delta^2 \zeta^i = (0, 0, \delta^2 \zeta^\rho)$ 
is parallel to the propagation surface of the wave, i.e. the equatorial plane.}
Using $\chi$, we find that the force-free condition is recovered 
with the second-order perturbation.
For simplicity, we assume $\partial_\Psi \chi = \partial^2_\Psi \chi = 0$ without any contradiction.

From Eq. (\ref{nodaeq04w}), the time evolution equation of $\chi$ is obtained as
\begin{equation}
\frac{1}{\sqrt{-g}} \partial_\lambda
(\sqrt{-g} g^{\Psi\Psi} g^{\lambda\beta} \partial_\beta \chi)
= s,
\label{eqchi4dlb_tx}
\end{equation}
where
\begin{equation}
s \equiv \frac{1}{\sqrt{-g}} \partial_\lambda (\sqrt{-g} Y^{\lambda\alpha\mu} \partial_\alpha \psi
\partial_\mu \psi ) 
= \frac{1}{\sqrt{-g}} [
\partial_T \Theta \partial_X \psi -\partial_X \Theta \partial_T \psi ]
\label{chieq}
\end{equation}
with $Y^{\lambda\alpha\mu} \equiv W^{\lambda\alpha\mu\nu} \partial_\nu \bar{\phi}_2 =
W^{\lambda\alpha\mu\rho}-\Omega_{\rm F}W^{\lambda\alpha\mu T}$ and
$\displaystyle \Theta \equiv \sqrt{-g} Y^{T\alpha X} \partial_\alpha \psi
= - \frac{I X^2}{\alpha^2} \partial_T \psi - R^2 (\Omega_{\rm F}-\Omega)
\partial_X \psi$ (see the derivation in Appendix \ref{appendb}).
The right-hand side of Eq. (\ref{eqchi4dlb_tx}), $s$, is recognized as the
source term of the wave equation of $\chi$ as the second-order
perturbation $\sim \psi^2$. Therefore, $\chi$ in Eq. (\ref{eqchi4dlb_tx}) 
represents the wave induced by the Alfv\'{e}n wave.
Introducing $\chi$, we confirm that the force-free condition holds.
The linear inhomogeneous equation (Eq. (\ref{eqchi4dlb_tx}))
shows the wave described by $\chi$
propagates isotropically in space as
by a fast wave \citep{bellman06},
while the complex formalism of the homogeneous linear equation of $\psi$ 
(Eq. (\ref{eq4i2})) suggests
strong anisotropy of the wave described by $\psi$, 
which is a characteristic property of an Alfv\'{e}n wave.

\section{Energy and angular momentum conservation
\label{sec3}}
In this section, we consider the energy and angular momentum transport
of the Alfv\'{e}n wave and the fast mode (the second-order perturbation). 

The Maxwell equations yield the energy-momentum conservation law,
\begin{equation}
\nabla_\nu T^{\mu\nu} = - f_{\rm L}^\mu ,
\label{coneneanm2emf}
\end{equation}
where $\displaystyle T^{\mu\nu} = {F^{\mu}}_\sigma F^{\nu\sigma} 
- \frac{1}{4} g^{\mu\nu} F_{\lambda\kappa} F^{\lambda\kappa}$ 
is the energy-momentum tensor of the electromagnetic field
and $f_{\rm L}^\mu = J^\nu {F^\mu}_\nu$ is the 4-Lorentz force density.
Here, $f_{\rm L}^\mu = 0$ is the ``force-free condition".
For the background equilibrium, we can confirm 
the Lorentz force of the equilibrium vanishes because
the 4-current density of the equilibrium vanishes,
$\bar{f}_\mu^{\rm L} = 0$.
We can also confirm that the Lorentz force with respect to the first-order perturbation
of Alfv\'{e}n wave vanishes, $\delta f_\mu^{\rm L} = 0$.
It is noted that $\delta J^\rho = \delta J^\Psi = 0$.
As concluded in the previous study in BTZ space-time \citep{koide22},
we have to take into account the second-order contribution to the energy and angular momentum,
because the first order 
always vanishes as shown later.
Furthermore, as we consider only the first-order perturbation of the Alfv\'{e}n 
wave, the second order of the right-hand side of Eq. (\ref{coneneanm2emf}) 
never vanishes: $\delta^{1+1} f_{\rm L}^\mu \ne 0$.
When we consider the second-order perturbation that corresponds to the fast mode, 
the right-hand side of
Eq (\ref{coneneanm2emf}) vanishes: $\delta^2 f_{\rm L}^\mu = \delta^{1+1} f_{\rm L}^\mu 
+\delta^{0+2} f_{\rm L}^\mu =$0.

To investigate the energy and momentum transfer due to the Alfv\'{e}n wave,
we consider the perturbation of $T^{\mu\nu}$ at first or higher order.
When we assume Eqs. (\ref{phi1per2nd}) and (\ref{phi2per2nd}),
the first order of $T^{\mu\nu}$ is calculated as
\begin{equation}
\delta T^\mu_\nu = \delta F^{\mu\lambda} \bar{F}_{\nu\lambda}
+ \bar{F}^{\mu\lambda} \delta F_{\nu\lambda}
- \frac{1}{2} g^\mu_\nu \bar{F}^{\lambda\kappa} \delta F_{\lambda\kappa},
\label{deltmunu}
\end{equation}
for which a detailed derivation is provided in Appendix \ref{appenda}.
The second- and third-order terms on the right-hand side of Eq. (\ref{deltmunu}) vanish
because of $\partial_\Psi \psi = 0$.
The first term on the right-hand side of Eq. (\ref{deltmunu}) is
\begin{equation}
\delta F^{\mu\lambda} \bar{F}_{\Psi\lambda}
= \delta F^{\mu\rho} \bar{F}_{\Psi\rho} = - \delta F^{\mu\rho}
= - W^{\mu\lambda\rho\rho},
\end{equation}
which 
\textcolor{black}{
is non-vanishing when either $\mu = \Psi$ or $\nu = \Psi$ holds}.
\textcolor{black}{Thus, when neither $\mu = \Psi$ or $\nu = \Psi$}, we have
\begin{equation}
\delta T^\mu_\nu = 0.
\label{deltzero}
\end{equation}
Equation (\ref{deltzero}) 
shows that the second-order perturbation should be considered
when investigating the energy transport of the linear Alfv\'{e}n wave.
If the force-free condition of the second-order 
perturbation is broken, the energy and angular momentum of the Alfv\'{e}n wave
are not conserved.
To verify the conservation of energy and angular momentum, we evaluate
the force-free condition up to second order in
the energy density and angular momentum density of the Alfv\'{e}n wave.
%

For the second-order perturbation, we have
\begin{equation}
\delta^2 J^\mu = \frac{1}{\sqrt{-g}} \partial_\nu (\sqrt{-g} \delta^2 F^{\mu\nu})
= \frac{1}{\sqrt{-g}} \partial_\nu (\sqrt{-g} W^{\mu\Psi\nu\kappa} \partial_\kappa \chi).
\end{equation}
The non-zero component of $\delta^2 J^\mu$ is 
\begin{equation}
\delta^2 J^\Psi
= \frac{1}{\sqrt{-g}} \partial_\nu (\sqrt{-g} g^{\Psi\Psi} g^{\nu\kappa} \partial_\kappa \chi).
\end{equation}
Using Eq. (\ref{eqchi4dlb_tx}), we find that 
the inhomogeneous term of the second-order perturbation of 
the energy conservation law (\ref{coneneanm2emf}) is written as
\begin{eqnarray}
\delta^2 f_T &=& \delta^2 J^\nu \bar{F}_{T\nu} + \delta J^\nu \delta F_{T\nu}
=  \delta^2 J^\Psi \bar{F}_{T\Psi} + \delta J^\nu \delta F_{T\nu} \nonumber \\
& = & \frac{1}{\sqrt{-g}} \Omega_{\rm F}
[\partial_\nu (\sqrt{-g} g^{\Psi\Psi} g^{\nu\kappa} \partial_\kappa \chi)
+\partial_\nu(\sqrt{-g} Y^{\mu\lambda\nu} \partial_\lambda \psi \partial_\mu \psi)] = 0,
\label{eq4fastwave}
\end{eqnarray}
which represents the fast wave 
induced by the Alfv\'{e}n wave (the first-order perturbation).
Thus, we have shown the total energy of the Alfv\'{e}n wave
and fast wave is conserved when considering second-order perturbations.

\subsection{The energy and angular momentum conservation laws 
in natural coordinates}

When $\xi^\mu$ is the Killing vector, we have the conservation law
\begin{equation}
\frac{\partial}{\partial x^0} ( \xi_\nu T^{0 \nu} )
+ \frac{1}{\sqrt{-g}} \frac{\partial}{\partial x^i} (\sqrt{-g} \xi_\nu T^{i \nu})
= \frac{1}{\sqrt{-g}} \partial_\mu (\sqrt{-g} T^{\mu\nu}) 
= \nabla_\mu (\xi_\nu T^{\mu\nu}) = - \xi^\nu f^{\rm L}_\nu .
\end{equation}
When $\xi^\nu f^{\rm L}_\nu$ vanishes,
$\xi_\nu T^{0 \nu}$ represents the density of the conserved value
and $\xi_\nu T^{i \nu}$ represents the density of the conserved quantity flux.

For the time-like Killing vector $\xi^\mu_{(T)} = (1, 0, 0, 0)$ 
and the axial Killing vector $\xi^\mu_{(\rho)} = (0, 0, 0, 1)$ 
in magnetic natural coordinates,
we have the energy and angular momentum conservation laws 
{
\begin{eqnarray}
\displaystyle \frac{\partial S^T}{\partial T}
&+& \frac{1}{\sqrt{-g}} \frac{\partial}{\partial X} (\sqrt{-g} S^X) 
+ \frac{1}{\sqrt{-g}} \frac{\partial}{\partial \Psi} 
(\sqrt{-g} S^\Psi)= f_T^{\rm L} = 0, 
\label{eneconeqconaco} \\
\frac{\partial M^T}{\partial T}
&+& \frac{1}{\sqrt{-g}} \frac{\partial}{\partial X} (\sqrt{-g} M^X) 
+ \frac{1}{\sqrt{-g}} \frac{\partial}{\partial \Psi} (\sqrt{-g} M^\Psi) = -f_\rho^{\rm L}
=0,
\label{anmconeqconaco} 
\end{eqnarray}
where we assume axisymmetry and $z$-direction translation symmetry, and}
$S^\mu = - \xi^{(T)}_\nu T^{\mu\nu}$ and $M^\mu = \xi^{(\rho)}_\nu T^{\mu\nu}$ 
are the 4-energy flux and 4-angular momentum flux, respectively, up to second order.

We now consider the energy density and energy flux to determine the energy conservation law
on the equatorial plane around a black hole.
The Killing vector for the energy conservation law is
$\xi_{(T)}^{{\mu}}=(1, 0, 0, 0)$.
Thus, the 4-energy flux density is 
\begin{equation}
{S}^{{\mu}} = - \xi_{(T)}^{{\nu}} T^{{\mu}}_{{\nu}} = - T^{{\mu}}_T.
\end{equation}
The energy conservation law is expressed as
\begin{equation}
\nabla_{{\mu}} {S}^{{\mu}}
= \frac{1}{\sqrt{-g}} \partial_{\mu} 
(\sqrt{-g} {S}^{{\mu}})
= \xi_{(T)}^{{\nu}} {f}_{{\nu}}^{\rm L} 
= {f}_T^{\rm L}
=  0.
\end{equation}
Hence, we have the energy density of the equilibrium, first-, and
second-order perturbations with respect to the linear Alfv\'{e}n wave (with $\delta^{1+1}$)
and the second-order fast wave (with $\delta^{2+0}$),
\begin{eqnarray}
{\bar{S}}^T &=& g^{\Psi\Psi} (g^{\rho\rho} - \Omega_{\rm F}^2 g^{TT}), \\
\delta {S}^T &=& 0 , \\
\delta^2 {S}^T &=&
\delta^{1+1} {S}^T +  \delta^{2+0} {S}^T  ,
\label{d2st}
\end{eqnarray}
where
\begin{eqnarray}
\delta^{1+1} {S}^T 
& = & \frac{1}{\sqrt{-g}} \left [ \frac{\lambda}{2} (\partial_T \psi)^2
+ \frac{S}{2} (\partial_X \psi)^2
 + \frac{2}{{\alpha}^2} \Omega_{\rm F}
\partial_T \psi \partial_X \psi  \right ] , 
\label{d11st} \\
\delta^{2+0} {S}^T & = & \frac{I}{\sqrt{-g}} \partial_X \chi 
- \frac{1}{{\alpha}^2} \Omega_{\rm F}  \partial_T \chi.
\label{d20st}
\end{eqnarray}
In the numerical calculations, we use the above formulation
(Eqs. (\ref{d2st})-(\ref{d20st})) with the tortoise 
coordinates (see Appendix \ref{append_numeth}).
We can obtain the energy flux density of the equilibrium, first-, and
second-order perturbations of the linear Alfv\'{e}n wave,
\begin{eqnarray}
{\bar{S}}^X &=& I \Omega_{\rm F},\\
\delta {S}^X &=&  0, \\
\delta^2 {S}^X &=&
- \partial_T \psi Y^{X\lambda\rho} \partial_\lambda \psi
- W^{\Psi\Psi X\rho} \partial_T \chi + \Omega_{\rm F}
W^{\Psi\Psi XX} \partial_X \chi
=  \delta^{1+1} {S}^X +  \delta^{2+0} {S}^X  ,
\label{d2sr}
\\ \delta^2 S^\Psi & = & - \partial_\Psi \psi (Y^{\Psi\Psi\rho} \partial_T \psi
+ \Omega_{\rm F} Y^{\Psi\Psi X} \partial_X \psi) = 0,
\end{eqnarray}
where
\begin{eqnarray}
\delta^{1+1} {S}^X & = &
- \frac{1}{\Sigma} \partial_T \psi \left [ 
\left ( \alpha^2 + R^2\Omega (\Omega_{\rm F} - \Omega) \right ) \partial_X \psi
- \frac{\Sigma \Omega I}{\alpha^2} \partial_T \psi  \right ] , \\
\delta^{2+0} {S}^X & = &
- \frac{I}{\Sigma}  \partial_T \chi
+ \Omega_{\rm F} \frac{\Delta}{\Sigma X^2} \partial_X \chi . 
\label{d20sr} 
\end{eqnarray}
Eventually, we obtain the energy conservation law of the Alfv\'{e}n wave
and the induced fast mode,
\begin{equation}
\frac{\partial}{\partial T} \delta^2 e^\infty + \frac{1}{\sqrt{-g}}
\frac{\partial}{\partial X}  (\sqrt{-g} \delta^2 S_{\rm P} ) = 0,
\end{equation}
where $\delta^2 e^\infty = \delta^{1+1} e^\infty + \delta^{2+0} e^\infty$,
$\delta^2 S_{\rm P} = \delta^{1+1} S_{\rm P} + \delta^{2+0} S_{\rm P}$,
$\delta^{1+1} e^\infty = \delta^{1+1} S^T$,
$\delta^{2+0} e^\infty = \delta^{2+0} S^T$,
$\delta^{1+1} S_{\rm P} = \delta^{1+1} S^X$, and
$\delta^{2+0} S_{\rm P} = \delta^{2+0} S^X$.
In the numerical calculations, we use the formulation 
(Eqs. (\ref{d2sr})--(\ref{d20sr}))
with the tortoise coordinates (see Appendix \ref{append_numeth}).

Next, we discuss the conservation law of the angular momentum.
The axial Killing vector $\xi^\mu_{(\rho)} = (0, 0, 0, 1)$ yields
the 4-angular momentum flux density $M^\mu = \xi^\nu_{(\rho)} T^\mu_\nu = F^{\mu\lambda} F_{\rho\lambda}$ ($\mu \ne \rho$), and
we obtain the angular momentum conservation law in the nonrotating natural coordinates,
\begin{equation}
\nabla_\mu M^\mu = \frac{1}{\sqrt{-g}} \partial_\mu (\sqrt{-g} M^\mu)
= \xi^\mu_{(\rho)} f_\mu^{\rm L} = f_\rho^{\rm L}.
\label{cnsangmom}
\end{equation}
We evaluate the angular momentum and the Lorentz force density
up to the second order:
\begin{eqnarray}
\delta^2 M^\mu &=& \delta F^{\mu\lambda} \delta F_{\rho\lambda} 
+ \delta^2 F^{\mu\lambda} \bar{F}_{\rho\lambda}
+ \bar{F}^{\mu\lambda} \delta^2 F_{\rho\lambda}
= - Y^{\mu\kappa\lambda} \partial_\kappa \psi \partial_\lambda \psi
+ W^{\Psi\Psi\mu\kappa} \partial_\kappa \chi, \\
\delta^2 f_\rho^{\rm L} &=& \delta J^\mu \delta F_{\rho\mu}
+ \delta^2 J^\nu \bar{F}_{\rho\mu} + \bar{J}^\nu \delta^2 F_{\rho\mu}
= - \frac{1}{\sqrt{-g}} \partial_\nu (\sqrt{-g} Y^{\mu\lambda\nu}
\partial_\kappa \psi \partial_\lambda \psi)
- \delta^2 J^\Psi  \nonumber \\
&=& \frac{1}{\sqrt{-g}} \partial_\nu (\sqrt{-g} Y^{\nu\lambda\mu}
\partial_\lambda \psi \partial_\nu \psi)) - \frac{1}{\sqrt{-g}}
\partial_\nu (\sqrt{-g} W^{\Psi\Psi\nu\kappa} \partial_\kappa \chi).
\end{eqnarray}
Equation (\ref{cnsangmom}) yields
\begin{equation}
\delta^2 f_\rho^{\rm L} =
\frac{1}{\sqrt{-g}} \partial_\nu (\sqrt{-g} Y^{\nu\lambda\mu}
\partial_\lambda \psi \partial_\nu \psi)) - \frac{1}{\sqrt{-g}}
\partial_\nu (\sqrt{-g} W^{\Psi\Psi\nu\kappa} \partial_\kappa \chi)=0,
\label{dl2flrho}
\end{equation}
which is identical to the equation for $\chi$ (Eq. (\ref{eqchi4dlb_tx})).
When and only when $I$, $\Omega_{\rm F}$, and $a_\ast$ are all zero,
the first term of the central part of Eq. (\ref{dl2flrho}) vanishes
(see the detailed calculation in Eq. (\ref{appdx_djrhol})).
In the case of $I=0$, $\Omega_{\rm F}=0$, or $a_\ast=0$, the Alfv\'{e}n 
wave never induces the fast wave.\\
Eventually, we obtain the angular momentum conservation law with respect to the Alfv\'{e}n wave
and the induced fast mode,
\begin{equation}
\frac{\partial \delta^2 l_z}{\partial T} + \frac{1}{\sqrt{-g}}
\frac{\partial}{\partial X}  (\sqrt{-g} \delta^2 M^X ) = 0,
\end{equation}
where $\delta^2 l_z = \delta^{1+1} l_z + \delta^{2+0} l_z$,
$\delta^2 M^X = \delta^{1+1} M^X + \delta^{2+0} M^X$,
$\delta^{1+1} l_z = \delta^{1+1} M^T$, and
$\delta^{2+0} l_z = \delta^{2+0} M^T$.

\subsection{The energy conservation law in the corotating natural frame}

We additionally consider the corotating natural coordinates $(T', X', \Psi', \rho')$
as in \citet{noda22},
\begin{eqnarray}
T' &=& T + \int \frac{I X^2 R^2}{\Gamma \Delta} (\Omega - \Omega_{\rm F}) dX, \\
X' &=& X, \\
\Psi' &=& \Psi , \\
\rho' &=& \rho - \Omega_{\rm F} T.
\end{eqnarray}
The time-like Killing vector of the corotating 
coordinates is given as $\xi_{(T')}^{\mu'} = (1, 0, 0, 0)$, and we have
\begin{equation}
\xi_{(T')}^\mu = \frac{\partial x^\mu}{\partial x^{\nu'}} \xi_{(T')}^{\nu'}
= (1, 0, 0, \Omega_{\rm F}).
\end{equation}
Then, the 4-vector of the energy flux density in the corotating coordinates is given
by
\begin{equation}
S^\mu_{(T')} = - \xi_{(T')}^\nu  T^\mu_\nu = S^\mu - \Omega_{\rm F} M^\mu.
\label{smutp}
\end{equation}
The conservation law of the Alfv\'{e}n wave for (1+1)-order perturbations is
\begin{equation}
\nabla_\mu \delta^{1+1} S^\mu_{(T')} 
= \frac{1}{\sqrt{-g}} \partial_\mu (\sqrt{-g} \delta^{1+1} S^\mu_{(T')})
= \delta^{1+1} (\xi_{(T')}^\mu  f_\mu) .
\label{enecon2cor}
\end{equation}
The right-hand side of Eq. (\ref{enecon2cor}) vanishes because
\begin{equation}
\delta^{1+1} (\xi_{(T')}^\mu  f_\mu) = 
\delta^{1+1} f_T + \Omega_{\rm F} \delta^{1+1} f_\rho
= \frac{1}{\sqrt{-g}} [\partial_\nu (-Z^{\nu\lambda} \partial_\lambda \psi)]
(\partial_T \psi + \Omega_{\rm F} \partial_\rho \psi) = 0.
\end{equation}
Eventually, we obtain the energy conservation law 
with respect to the Alfv\'{e}n wave in the corotating magnetic natural frame:
\begin{eqnarray}
&& \frac{1}{\sqrt{-g}} \partial_\mu (\sqrt{-g} \delta^{1+1} S^\mu_{(T')})
= \frac{1}{\sqrt{-g}} \partial_\mu 
(\sqrt{-g} (\delta^{1+1}S^\mu - \Omega_{\rm F} \delta^{1+1} M^\mu)) \nonumber \\
&=& \frac{\partial}{\partial T} (\delta^{1+1}S^T - \Omega_{\rm F} \delta^{1+1} M^T)
+ \frac{1}{\sqrt{-g}} \frac{\partial}{\partial X} [\sqrt{-g}(\delta^{1+1}S^X - \Omega_{\rm F} \delta^{1+1} M^X)] \nonumber \\
&+& \frac{1}{\sqrt{-g}} \frac{\partial}{\partial \Psi} [\sqrt{-g}(\delta^{1+1}S^X - \Omega_{\rm F} \delta^{1+1} M^\Psi)] \nonumber \\
&=& \frac{\partial}{\partial T} \left ( \delta^{1+1}S^T - \Omega_{\rm F} \delta^{1+1} M^T
+ \frac{1}{2} Z^{\Psi\Psi} \psi^2 \right )
+ \frac{1}{\sqrt{-g}} \frac{\partial}{\partial X} [\sqrt{-g}(\delta^{1+1}S^X - \Omega_{\rm F} \delta^{1+1} M^X)] \nonumber \\
&=& \frac{1}{\sqrt{-g}} \left \{ \frac{\partial}{\partial T} \left [ 
\frac{\lambda}{2} \left ( \frac{\partial \psi}{\partial T} \right )^2
+ \frac{S}{2} \left ( \frac{\partial \psi}{\partial X} \right )^2
+ \frac{K}{2} \psi^2 \right ]
+ \frac{\partial}{\partial X} \left [ 
-S \frac{\partial \psi}{\partial X} \frac{\partial \psi}{\partial T}
+ V \left ( \frac{\partial \psi}{\partial T} \right )^2 \right ] \right \} = 0.
\end{eqnarray}
We write the energy conservation law simply as
\begin{equation}
\frac{\partial}{\partial T} \delta^2 e^{\infty\prime}
+ \frac{1}{\sqrt{-g}} \frac{\partial}{\partial X} (\sqrt{-g} \delta^2 S_{\rm P}^{\prime})
=0,
\label{conslawenecorot}
\end{equation}
where 
\begin{eqnarray}
\displaystyle \delta^2 e^{\infty\prime} &=& \frac{1}{\sqrt{-g}}
\left [ \frac{\lambda}{2} \left ( \frac{\partial \psi}{\partial T} \right )^2
+ \frac{S}{2} \left ( \frac{\partial \psi}{\partial X} \right )^2
+ \frac{K}{2} \psi^2 \right ], \\
\displaystyle \delta^2 S_{\rm P}^{\prime} &=& \frac{1}{\sqrt{-g}}
\left [ - S \frac{\partial \psi}{\partial T} \frac{\partial \psi}{\partial X}
+ V \left ( \frac{\partial \psi}{\partial T} \right )^2 \right ]
\end{eqnarray}
are the energy density
and energy flux of the Alfv\'{e}n wave observed by the corotating
magnetic natural frame, respectively.
Then, in the corotating frame, the law of conservation of energy holds true
even only with the Alfv\'{e}n wave.

For the angular momentum in the corotating frame, we have
\begin{eqnarray}
M^\nu_{(\rho')} = \xi^\mu_{(\rho')} T^\nu_\mu = \xi^\rho_{(\rho')} 
= T^\nu_\rho = M^\nu,
\end{eqnarray}
because
\begin{eqnarray}
\xi^\mu_{(\rho')} = \partial_{\rho'} 
= \frac{\partial x^\mu}{\partial x^{\nu'}} \xi^{\nu'}_{(\rho')} 
= \frac{\partial x^\mu}{\partial x^{\rho'}}
= \left (0, 0, 0, \frac{\partial \rho}{\partial \rho'} \right )
= (0, 0, 0, 1).
\end{eqnarray}
We then find that the angular momentum densities in the corotating frame
($x^{\mu^\prime}$)
and the non-rotating frame ($x^{\mu}$) are identical.
This means that the angular momentum of the Alfv\'{e}n wave alone, as described by $\psi$,
is not conserved 
even in the corotating frame, while the energy of the Alfv\'{e}n wave alone is conserved
in the corotating frame. To conserve the angular momentum, we have to
consider the fast mode described by the second-order perturbation $\chi$.

It is noted that Eq. (\ref{smutp}) yields
\begin{equation}
\delta^{1+1} e^\infty =
\delta^{1+1} e^{\infty\prime} + \Omega_{\rm F} \delta^{1+1} l^{z'},
\end{equation}
where $\delta^{1+1} e^\infty = \delta^{1+1} S^T$,
$\delta^{1+1} e^{\infty\prime} = \delta^{1+1} S^{T'}_{(T')}$, and
$\delta^{1+1} l^{z'} = \delta^{1+1} M^{T'}_{(\rho')} = \delta^{1+1} M^{T}$.
Here, $\Omega_{\rm F} \delta^{1+1} l^{z'}$ corresponds to the rotational energy
and is converted to the energy of the fast mode because $\delta^{1+1} e^{\infty\prime}$ 
is conserved from Eq. (\ref{conslawenecorot}).

\section{Numerical method
\label{secmet}}

To perform 1D numerical simulations of the force-free field using Eqs. (\ref{newtontoyeq})
and (\ref{eqchi4dlb_tx}), 
we use the multi-dimensional two-step Lax-Wendroff scheme for 
$\VEC{u} = (\psi, \xi, \zeta, \chi, \upsilon, \sigma)$ whose details
are provided in Appendix \ref{appendnum}.
Here, $\xi, \zeta, \upsilon$, and $\sigma$ 
are the new variables introduced for the numerical calculation.
We utilize the tortoise coordinate $\displaystyle x = X - 2M +
r_{\rm H} \log (|X -r_{\rm H}|/(2M - r_{\rm H}))$
(see the details in Appendix \ref{appendnum}).
In the numerical calculations, we normalize the length using $M$. 
This scheme sometimes causes numerically artificial structures due to
the Gibbs phenomena. To avoid such numerical phenomenon, we employ
a smooth profile for the initial variables as follows.
%
The initial condition of an outwardly/inwardly propagating single pulse of a field is given as:
\begin{eqnarray}
\psi(x) &\equiv& \left \{ 
\begin{array}{cc} \displaystyle \left [ 1 - \left | \frac{x-x_0}{w/2} \right |^{2m} \right ]^n
& \displaystyle \left (x_0 - \frac{w}{2} \leq x \leq x_0 + \frac{w}{2} \right ) \\
0 & ({\rm other}) \end{array} \right . , \nonumber \\
\xi(x) &=& \bar{v}_{\rm ph}^\pm(x) \psi(x) , \\
\zeta(x) &=& 0 , \nonumber \\
\chi(x) &=& 0  , \nonumber \\
\upsilon(x) &=& 0 , \nonumber \\
\sigma(x) &=& 0 , \nonumber
\end{eqnarray}
where the plus and minus signs of
$\displaystyle \bar{v}_{\rm ph}^\pm (x) = \frac{dx}{dX} v_{\rm ph}^\pm (x)$ 
in the equation with respect to 
$\xi(x)$ are taken for the outwardly and
inwardly propagating waves, respectively, 
and $v_{\rm ph}^\pm$ is given by Eq. (\ref{vphpm}).
Note that $x_0$ and $w$ give the center and the width of the pulse of the Alfv\'{e}n wave and $m$ and $n$ 
are constants of the pulse shape. In this paper, we set
$n=32$ and $m=\log(1-2^{-1/32})/\log(1-2^{-1/4})$.

As shown in the next section, we have very smooth numerical results
without numerically artificial structures.
In this paper, we set the width and position of the outwardly 
and inwardly propagating pulses to be $w=0.5$, $x_0=-1.6$ (outward pulse)
and $w=3$, $x_0=2$ (inward pulse).

At the inner and outer boundaries, $x=x_{\rm min}$ and $x=x_{\rm max}$,
the free boundary conditions $\VEC{u}_0 - \VEC{u}_1 = \VEC{0}$,
$\VEC{u}_{I} - \VEC{u}_{I-1} = \VEC{0}$
are used to mimic the radial boundary condition,
where $\VEC{u}_{0}$ and $\VEC{u}_{I}$ are the variables
at the boundary and $\VEC{u}_{1}$ and $\VEC{u}_{I-1}$ are
the values in the neighborhood.

\section{Numerical results 
\label{secres}}

\begin{table}[b]
\begin{center}
\begin{tabular}{ccccccccc} \hline \hline
\begin{minipage}[h]{6em} Propagation direction \end{minipage} 
& \begin{minipage}[h]{10em} \begin{center} 
Relationship between \\ $\Omega_{\rm F}$ and $\Omega_{\rm H}$ 
\end{center} \end{minipage}  &
$\Omega_{\rm F}$ & $\Omega_{\rm H}$ & $a$ &
\begin{minipage}[h]{8em} \begin{center} Zeroth order\\ energy flux
\end{center} \end{minipage} & $x_0$ & $w$ 
& \begin{minipage}[h]{5em}  \begin{center}  Figure \\ number  \end{center} \end{minipage}
  \\ \hline
 Outward & $0<\Omega_{\rm F}<\Omega_{\rm H}$ & 0.027 & 0.0505 & 0.2 &
$\bar{S}^X > 0 $ & 5.0 & 8.0 & \ref{fig_out1000p6}, \ref{fig_outh12}  \\ \hline 
 Outward & $\Omega_{\rm F} \ge \Omega_{\rm H}$ & 0.06 & 0.0505 & 0.2 &
$\bar{S}^X < 0 $ & 5.0 & 8.0 & \ref{fig_bzofp6}, \ref{fig_bzfh12}   \\ \hline 
 Inward & $0<\Omega_{\rm F}<\Omega_{\rm H}$ & 0.027 & 0.0505 & 0.2 & 
$\bar{S}^X > 0 $ & 20.0 & 20.0 & \ref{fig_refp6}, \ref{fig_refh12} \\ \hline 
\end{tabular}
\end{center}
\caption{1D force-free magnetodynamics (FFMD) simulations of an Alfv\'{e}n wave along
a magnetic field line
at the equatorial plane around a Kerr black hole. \label{tabl1}}
\end{table}

We performed 1D numerical simulations of Alfv\'{e}n waves
along a stationary magnetic field line with $\Omega_{\rm F}=0.027$
($0 < \Omega_{\rm F} < \Omega_{\rm H} = 0.0505$)
and $\Omega_{\rm F}=0.06$ ($\Omega_{\rm F} > \Omega_{\rm H} = 0.0505$) at $\Psi=0$ 
around a black hole with spin parameter $a=0.2$ 
using Eqs. (\ref{newtontoyeq}) and (\ref{chieq})
{(Table \ref{tabl1})}.
Here, we set $M$ to unity.
In this case, we have the horizon radius $r_{\rm H}=1+\sqrt{1-a_\ast^2}=1.9798$.
The radii of the inner and outer light surfaces $r_{\rm LS}^-=1.98$
($x_{\rm LS}^-= -3.03$) and $r_{\rm LS}^+=36.0$ ($x_{\rm LS}^+= 48.7$)
in the case of $\Omega_{\rm F}=0.027$, $r_{\rm LS}^-=1.98$ ($x_{\rm LS}^-= -57.8$) and  
$r_{\rm LS}^+=18.7$ ($x_{\rm LS}^+= 30.0$)
in the case of $\Omega_{\rm F}=\Omega_{\rm H}=0.0505$ and $r_{\rm LS}^-=1.98$ 
($x_{\rm LS}^-= -6.55$) and  $r_{\rm LS}^+=15.6$ ($x_{\rm LS}^+= 26.5$)
in the case of $\Omega_{\rm F}=0.06$. 
{The ergosphere is the region inside the static limit surface, $r \le r_{\rm ergo} = 2$.}

The unstable radial ranges of the Alfv\'{e}n wave discussed in Section \ref{disrelalfwvinst} 
for the cases with
$\Omega_{\rm F} = 0.027$ and $\Omega_{\rm F} = 0.06$ are $r > r_{\rm m}=76.8$
($x > x_{\rm m} = 91.1$) and $r > r_{\rm m}=15.8$ and ($x > x_{\rm m} = 26.7$),
respectively. The numerical calculations in this paper are performed
in the stable radial range ($r < r_{\rm m}$) except for the region
$26.7 = r_{\rm m} < r < 30$ in the case of $\Omega_{\rm F} = 0.06 > \Omega_{\rm H}$.
In the case of $\Omega_{\rm F} > \Omega_{\rm H}$, the Alfv\'{e}n wave
does not enter the unstable region as shown in Section \ref{caseoutprobzoff}. 
However, the Alfv\'{e}n mode
is amplified exponentially when the perturbation is added in the unstable
region ($r > r_{\rm m}$). In this paper, we do not perform numerical simulations
for the unstable Alfv\'{e}n mode.

The initial $\psi$ conditions are given by $w=0.5$ and $x_0 = -1.5$ at $T=0$ in the case
of an outwardly propagating pulse in background magnetic fields with
$\Omega_{\rm F} = 0.027$, $\Omega_{\rm F} = 0.0505$, and $\Omega_{\rm F} = 0.06$.
In the case of an inwardly propagating pulse, we set $w=3$, $x_0=4$, 
and $\Omega_{\rm F} = 0.027$ for the initial condition of $\psi$ at $T=0$.
The initial condition of $\chi$ is given by $\chi = 0$ in all cases at $T=0$.

\subsection{An outwardly propagating Alfv\'{e}n wave in the case of $\Omega_{\rm F} < 
\Omega_{\rm H}$}

\begin{figure}[H]
\begin{center}
\includegraphics[width=17cm]{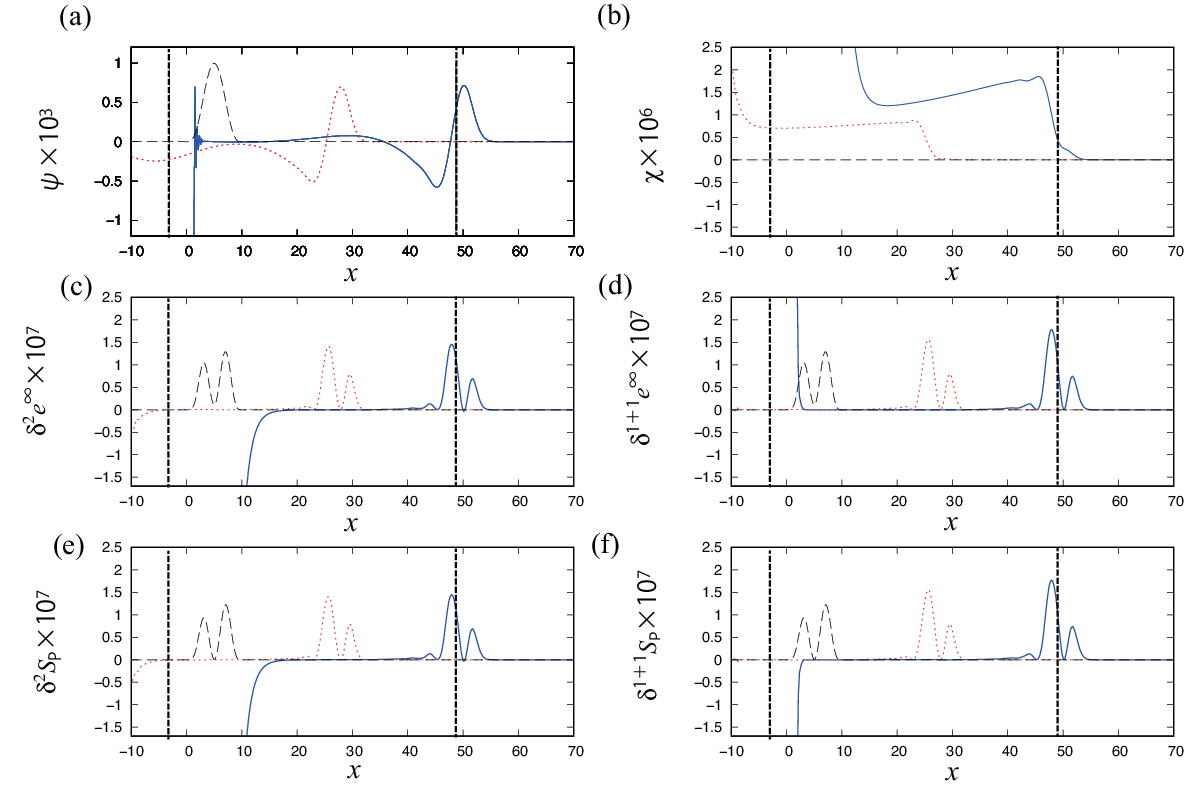}
\end{center}
\vspace{1cm}
\caption{The time evolution of the variables of a force-free electromagnetic
field for an initially outwardly propagating pulse of an Alfv\'{e}n wave 
along a magnetic field line with $\Omega_{\rm F} =0.027$
(corresponding to the case of black hole energy extraction by the Blandford--Znajek mechanism,
$0 < \Omega_{\rm F} < \Omega_{\rm H}$)
around a Kerr black hole with spin parameter $a=0.2$ at
$T=$ 0.0 (black dashed line), 25.0 (red dotted line), 50.0 (blue, thick solid line).
The dashed vertical lines indicate the inner and outer light surfaces.
\label{fig_out1000p6}}
\end{figure}

\begin{figure}[H]
\begin{center}
\includegraphics[width=17cm]{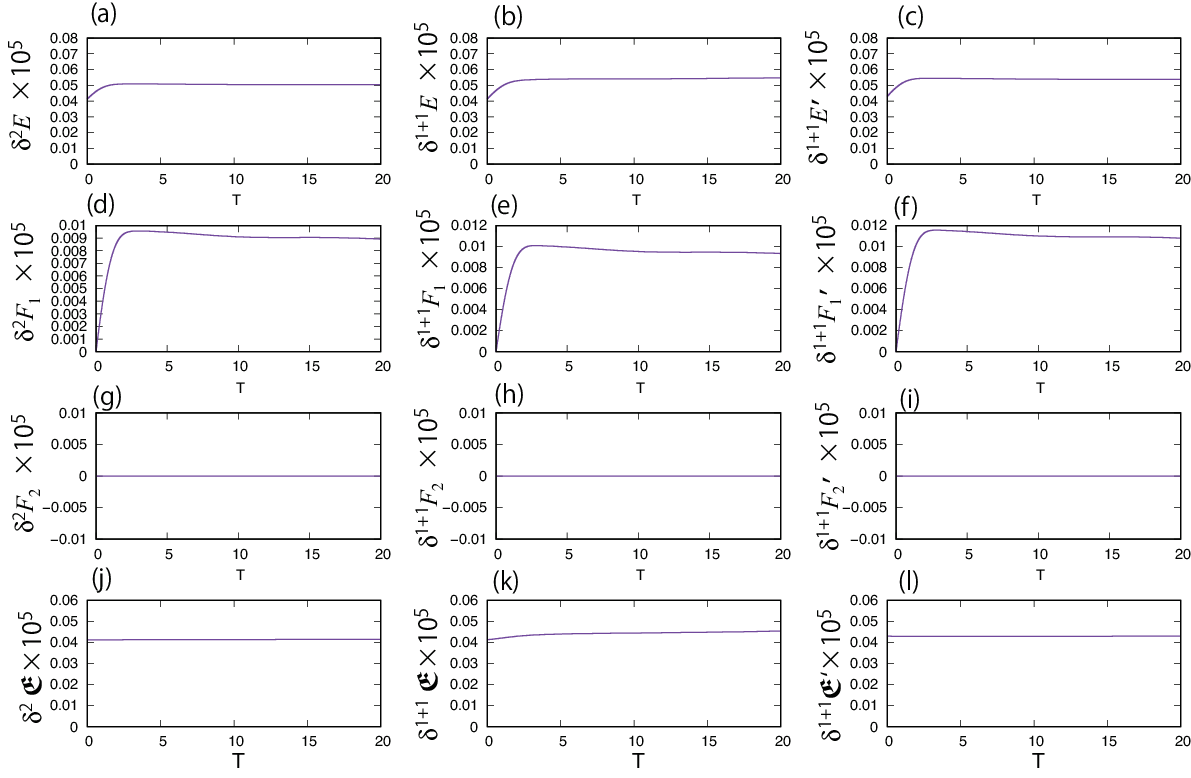}
\end{center}
\vspace{1cm}
\caption{The time evolution of the energy balance of the Alfv\'{e}n 
and induced fast waves.
Top panels: The wave energies. Second panels: The energy fluxes at the left edge.
Third panels: The energy fluxes at the right edge. Bottom panels: The balance of energy.
Left panels: The Alfv\'{e}n and fast waves. Middle panels: The Alfv\'{e}n wave alone.
Right panels: The Alfv\'{e}n wave observed in the corotating frame.
\label{fig_outh12}}
\end{figure}

Figure \ref{fig_out1000p6} shows that the Alfv\'{e}n wave propagates outward and that the
fast wave is caused by the Alfv\'{e}n wave
and propagates toward both the outward and inward sides 
in the case of $\Omega_{\rm F} = 0.027$,
$a=0.2$, $w=0.5$, and $x_0=-1.5$ (Fig. \ref{fig_out1000p6} (a), (b)). 
The lines in each panel show the results at $T=0$ (black dashed line),
$T=17.5$ (red dotted line), and $T=35.0$ (blue, thick solid line) 
(Fig. \ref{fig_out1000p6} (d)).
The energy density of the Alfv\'{e}n wave
$\sqrt{-g} \delta^{1+1} e^\infty$
is initially positive and remains positive, where
the energy density of the Alfv\'{e}n wave
$\sqrt{-g} \delta^{1+1} e^\infty$
is calculated from only $\delta^{1+1} e^\infty$ of $\psi$ without $\chi$
(see the details in Appendix \ref{appendetbtz}). 
The total energy density of the Alfv\'{e}n wave
and the induced fast wave, $\sqrt{-g} \delta^{2} e^\infty$,
remains predominantly positive (Fig. \ref{fig_out1000p6} (c)). 
The integrated energy over
the calculation region ($E(t)$ in Eq. (\ref{eqofet})) is shown in 
Fig. \ref{fig_outh12} (i). 
Here, we calculate the balance with respect to the energy numerically,
$\delta^2 \mathfrak{E} = \delta^2 E(t) + \delta^2 F_1(t) - \delta^2 F_2(t)$,
which should be a constant $\delta^2 E(0)$.
$\delta^{1+1} \mathfrak{E}$ represents the numerical energy balance for the Alfv\'{e}n wave
and $\delta^{1+1} \mathfrak{E}'$ represents the numerical energy balance for the Alfv\'{e}n wave
in the corotating natural frame.
The total energy balance of the Alfv\'{e}n
wave and the induced wave remains constant, which means it
is conserved.
The induction of the fast wave by the Alfv\'{e}n wave is explained by the relativistic mechanism, where
the Alfv\'{e}n wave has the angular momentum as in Section
\ref{sec_peralfwindw} (Fig. \ref{pontie_indwav}).


\subsection{An outwardly propagating Alfv\'{e}n wave in the case of
$\Omega_{\rm F} >\Omega_{\rm H}$
\label{caseoutprobzoff}}

\begin{figure}[H]
\begin{center}
\includegraphics[width=17cm]{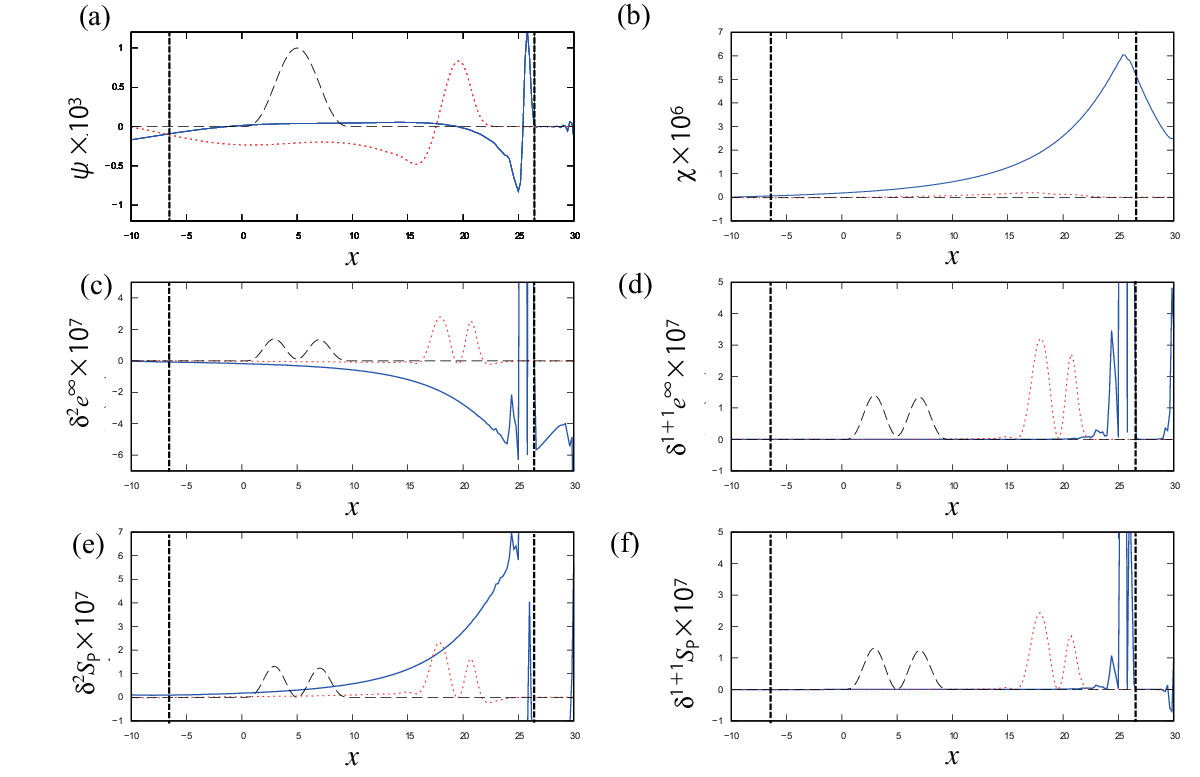}
\end{center}
\vspace{1cm}
\caption{Similar to Fig. \ref{fig_out1000p6}, but for 
the case of an initially outwardly propagating pulse 
with a background magnetic field $\Omega_{\rm F} = 0.06$
($\Omega_{\rm F} > \Omega_{\rm H}$: corresponding to the case with no outward power from the Blandford--Znajek
mechanism)
at $T=$ 0.0 (black dashed line), 20.0 (red dotted line), 40.0 (blue, thick solid line).
\label{fig_bzofp6}}
\end{figure}

As shown in Fig. \ref{fig_bzofp6}, 
we also performed a calculation for an Alfv\'{e}n wave along a magnetic field
line with $\Omega_{\rm F} =0.06$, where the rotational energy of the black hole
is not extracted by the Blandford--Znajek mechanism ($\Omega_{\rm F} > \Omega_{\rm H}
= 0.0505$).
The Alfv\'{e}n wave propagates outward and the fast wave is induced.
The induced wave propagates toward both sides as the case of $\Omega_{\rm F}= 0.027$
(where the Blandford--Znajek mechanism extracts the black hole rotational energy). 
The width of the pulse of the Alfv\'{e}n wave decreases until it vanishes and
cannot pass outward through the outer light surface 
($X=r_{\rm LS+}=18.7$,$x=x_{\rm LS+}=30.0$). 
Meanwhile, the fast wave induced by the Alfv\'{e}n wave passes 
through the outer light surface smoothly.

\begin{figure}[H]
\begin{center}
\includegraphics[scale=0.6]{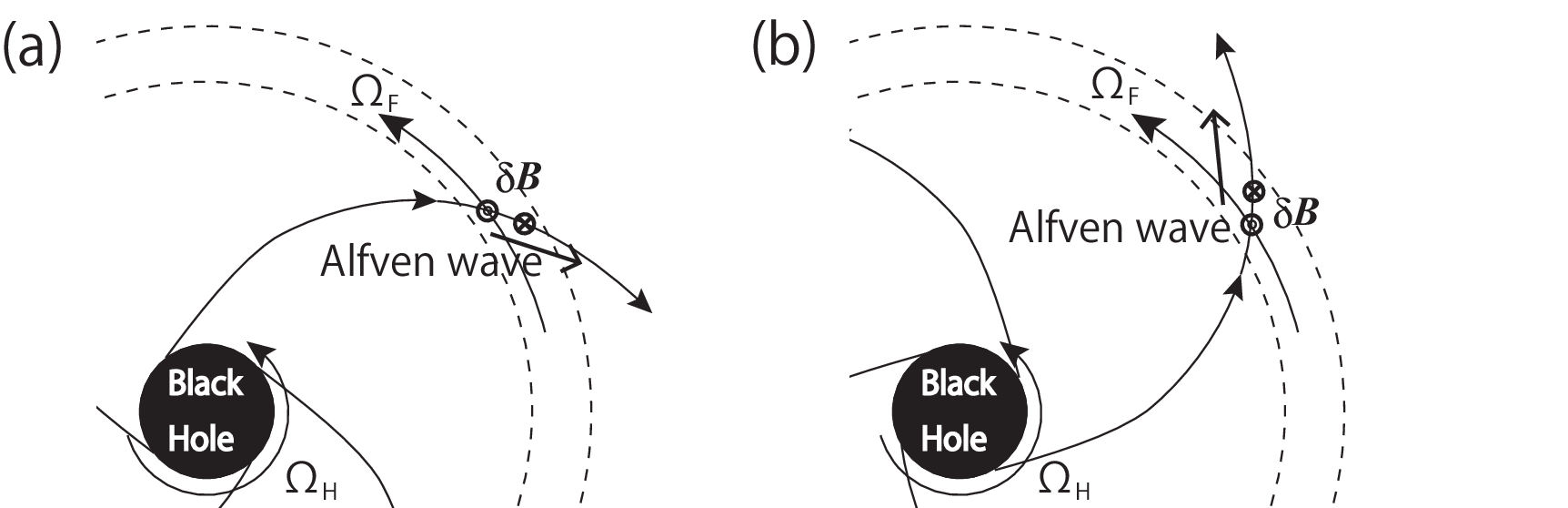}
\end{center}
\caption{A schematic of Alfv\'{e}n wave pulse propagation along the
field line of a force-free magnetic field
around a spinning black hole (a) in the case of $0 <\Omega_{\rm F} < \Omega_{\rm H}$
and (b) in the case of $\Omega_{\rm F} \ge \Omega_{\rm H}$.
\label{pontie_zentai2}}
\end{figure}

The slowing of the Alfv\'{e}n wave near the outer light surface is explained
by the dispersion relation of the Alfv\'{e}n wave provided in Appendix \ref{appdx_disprel}.
This slowing originates from the shape of the magnetic field line
(see Fig. \ref{pontie_zentai2}). 
The azimuthal component of the magnetic field in the Boyer--Lindquist coordinates 
is given by 
$\displaystyle \bar{B}^\phi = ^\ast \hspace{-1pt} \bar{F}^{t\phi}=-\frac{I}{\Delta} = -\frac{2M}{\Delta r_{\rm H}} 
(\Omega_{\rm H} - \Omega_{\rm F})$, while the radial component of the magnetic field
is given by $\displaystyle \bar{B}^r = ^\ast \hspace{-1pt} \bar{F}^{tr}=\frac{1}{\Sigma} > 0$ 
in the equatorial plane $\displaystyle \left ( \theta = \frac{\pi}{2} \right )$.
When $0 < \Omega_{\rm F} < \Omega_{\rm H}$, the magnetic field line bends
in the opposite direction to the rotation of the magnetic field line outward,
and the azimuthal component of the velocity of the Alfv\'{e}n wave
never exceeds the speed of light (Fig. \ref{pontie_zentai2}(a)). Otherwise, the magnetic field line bends
in the opposite direction, and the wave velocity exceeds the speed of light if it passes
over the outer light surface (Fig. \ref{pontie_zentai2}(b)).
Such a situation never occurs, and then the Alfv\'{e}n wave slows down to stop before the
outer light surface, as confirmed by the numerical simulation
shown in Fig. \ref{fig_bzofp6}.

\begin{figure}[H]
\begin{center}
\includegraphics[width=17cm]{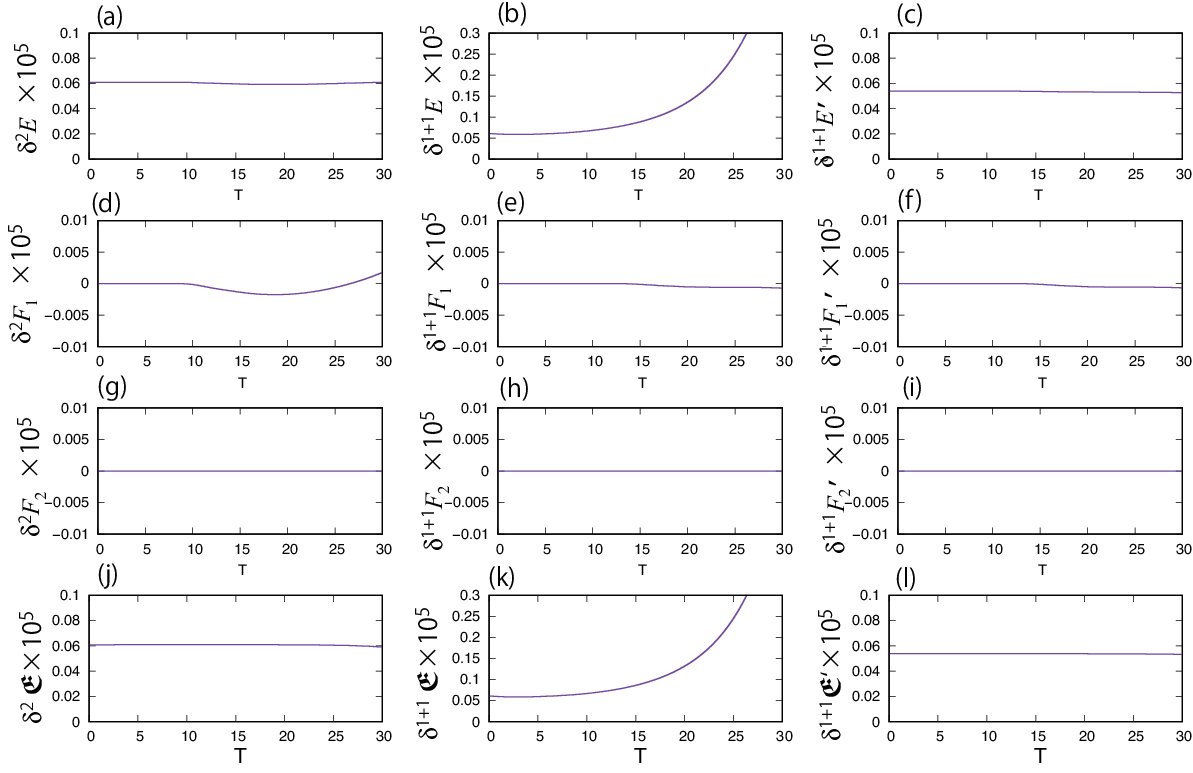}
\end{center}
\vspace{1cm}
\caption{Similar to Fig. \ref{fig_outh12}, but for 
the case of an initially outwardly propagating pulse 
with the background magnetic field $\Omega_{\rm F} = 0.06$
($\Omega_{\rm F} > \Omega_{\rm H}$).
\label{fig_bzfh12}}
\end{figure}

\subsection{An inwardly propagating Alfv\'{e}n wave}

\begin{figure}[H]
\begin{center}
\includegraphics[width=17cm]{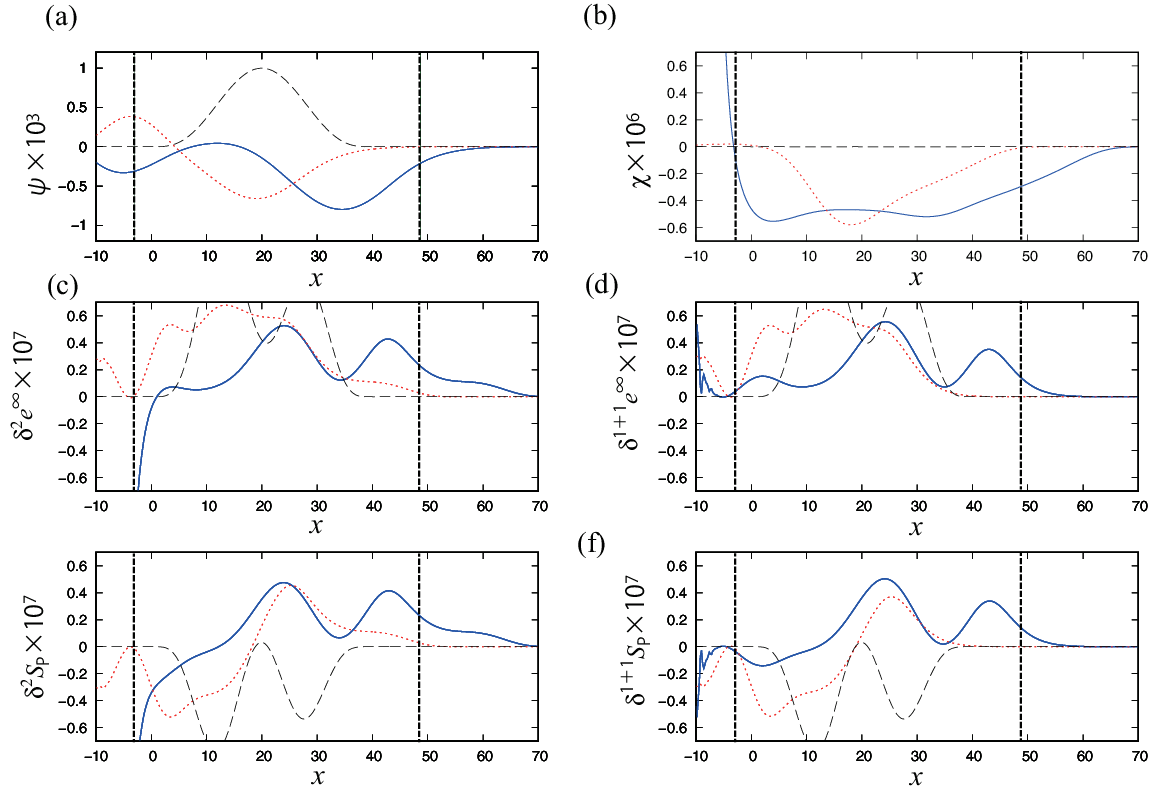}
\end{center}
\vspace{1cm}
\caption{The time evolution of the variables of a force-free electromagnetic
field for an initially inwardly propagating pulse of an Alfv\'{e}n wave from a region
in the ergosphere along a magnetic field line with $\Omega_{\rm F} =0.027$
(corresponding to the case of black hole energy extraction via the Blandford--Znajek mechanism,
$0 < \Omega_{\rm F} < \Omega_{\rm H}$)
around a spinning black hole with spin parameter $a=0.2$ at
$T=$ 0.0 (black dashed line), 17.5 (red dotted line), 35.0 (blue, thick solid line).
The dashed-dotted lines show the location of the light surfaces.
The reflection of the wave is shown around $x=5$.
\label{fig_refp6}}
\end{figure}

Figure \ref{fig_refp6} shows the numerical result for the inwardly propagating
pulse of an
Alfv\'{e}n wave along a magnetic field line with $\Omega_{\rm F}=0.027$ around
a spinning black hole with $a=0.2$.
The Alfv\'{e}n wave pulse propagates inward and induces the fast wave.
The induced wave propagates mainly inward with the Alfv\'{e}n wave pulse.
The pulse of the Alfv\'{e}n wave is reflected as described by \citet{noda22}
around the point $x=5$, and the pulse is separated into two pulses.
This reflection can be explained through the analogy of a wave along a string 
with an inhomogeneous mass line density $\lambda$, tension $S$, and 
spring constant $K$, as described by Eq. (\ref{newtontoyeq}).
However, the reflection rate of the wave is less than unity and the superradiance
is not shown.

\begin{figure}[H]
\begin{center}
\includegraphics[width=17cm]{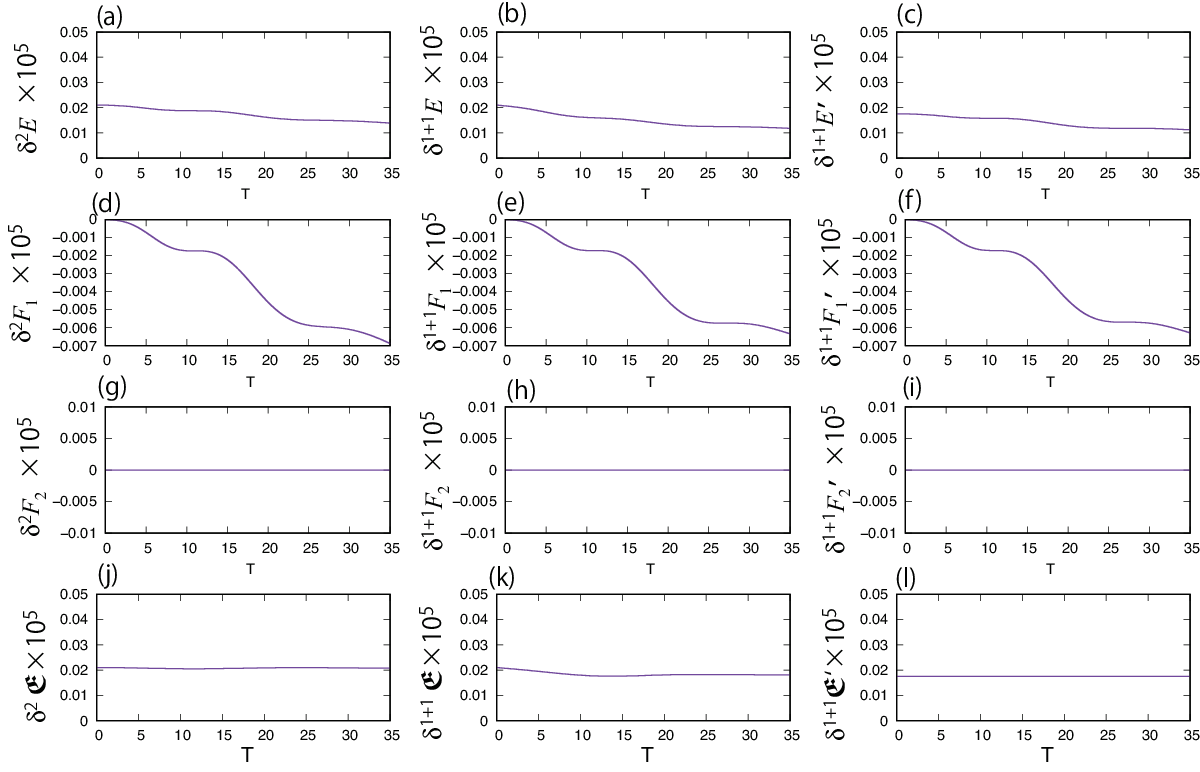}
\end{center}
\vspace{1cm}
\caption{Similar to Fig. \ref{fig_outh12}, but for 
the case of an initially inwardly propagating pulse 
with the background magnetic field $\Omega_{\rm F} = 0.027$
($0 < \Omega_{\rm F} < \Omega_{\rm H}$).
\label{fig_refh12}}
\end{figure}

The energy contribution of the induced wave is not so large
that $\sqrt{-g} \delta^2 e^\infty$ is almost equal to 
$\sqrt{-g} \delta^{1+1} e^\infty$.
In fact, the time evolution of the total energy almost identical to
that of the Alfv\'{e}n wave (Fig. \ref{fig_refh12}). 
Thus, in the case of an inward-propagating wave from the outer region,
the fast induced wave is negligible in the energy conservation law.

\section{Concluding remarks
\label{secdis}}

In this work, we have performed 1D linear analysis and numerical simulations of force-free 
Alfv\'{e}n waves and induced fast waves along a magnetic
field line at the equatorial plane 
around a black hole in a non-rotating natural 
coordinate frame using Eq. (\ref{newtontoyeq}) for $\psi$ 
and Eq. (\ref{chieq}) for $\chi$
to satisfy the force-free condition and energy conservation.
We identified the following interesting phenomena.
\begin{itemize}
\item The Alfv\'{e}n wave induces a fast magnetosonic wave when its angular momentum
changes due to the rotation and bending of the background
magnetic field lines.
The total energy and angular momentum
of the Alfv\'{e}n wave and the fast wave are conserved, while the energy and 
angular momentum of the Alfv\'{e}n wave alone are not conserved. 
\item In the case of $\Omega_{\rm F} > \Omega_{\rm H}$, the outwardly propagating
Alfv\'{e}n wave never passes through the outer light surface, while in the case of
$\Omega_{\rm F} < \Omega_{\rm H}$, it
passes through the outer light surface smoothly.
The fast wave passes through the outer light surface in any case.
\item The inwardly propagating Alfv\'{e}n wave is reflected around the region
near the inner light surface. Within the numerical calculations, 
we found no phenomenon suggesting
superradiance of the reflected Alfv\'{e}n wave.
\item The Alfv\'{e}n wave is unstable over the region $r = X > r_{\rm m}$,
where $r_{\rm m}$ is the solution for $K=0$ outside the outer light surface
when $\displaystyle \Omega_{\rm F} > \Omega_{\rm c} = \frac{2M}{2M+r_{\rm H}} |\Omega_{\rm H}|$.
When $\displaystyle \Omega_{\rm F} < \Omega_{\rm c}$, the Alfv\'{e}n wave
propagates stably.
\end{itemize}
The fast wave induced by the Alfv\'{e}n wave is attributed to a change in
the angular momentum of the Alfv\'{e}n wave (Fig. \ref{pontie_indwav}).
The Alfv\'{e}n wave has momentum and angular momentum in the relativistic.
framework. \footnote{In the special relativistic framework, 
the momentum density of the transverse wave is given by 
$\vec{S}/c^2$, where $\vec{S}$ is the
Poynting flux and $c$ is the speed of light.
In the nonrelativistic limit ($c \rightarrow \infty$), 
the momentum density vanishes.}
When the Alfv\'{e}n wave propagates along a curved
magnetic field line, the angular momentum changes due to 
the torque from the background magnetic field.
The reaction of the torque induces the fast wave described by $\chi$. 
The above mechanism is expected to work in the relativistic 
magnetohydrodynamic (MHD) framework.
Therefore, in a non-relativistic MHD framework,
the angular momentum of the Alfv\'{e}n wave vanishes, and the fast wave is not induced.
The second-order fast wave is induced by the Alfv\'{e}n wave only
in relativistic MHD.
This mechanism explains the conversion of the Alfv\'{e}n wave into a
fast magnetosonic wave in the pulsar magnetosphere, as shown by \citet{yuan21}.
The authors showed that an Alfv\'{e}n wave loses energy due to induction of the fast mode.
The same energy conversion of the Alfv\'{e}n wave is found in the case
with $\Omega_{\rm F} = 0.06 > \Omega_{\rm H}$ (Fig. \ref{fig_bzfh12}).
However, in the case with $\Omega_{\rm F} = 0.027 < \Omega_{\rm H}$ ,
the energy of the Alfv\'{e}n wave pulse increases as it propagates outward
from the ergosphere (Fig. \ref{fig_outh12}).

In the case of outward power radiation via the Blandford--Znajek mechanism
($0 < \Omega_{\rm F} <\Omega_{\rm F}$), the outwardly propagating
Alfv\'{e}n wave pulse passes through the outer light surface smoothly.
In the case without this stationary outward power radiation
($\Omega_{\rm F} > \Omega_{\rm F}$), the pulse
of the Alfv\'{e}n wave never passes the outer light surface, while the induced
fast waves pass smoothly through.
This is because the speed of the Alfv\'{e}n wave along the magnetic field line exceeds 
the speed of light when the outward Alfv\'{e}n wave has passed through the outer light 
surface because the magnetic field line is oblique with $B^\phi/B^r > 0$
in the case of $\Omega_{\rm F} > \Omega_{\rm H}$ (see Fig. \ref{pontie_zentai2}(b)).

The Alfv\'{e}n wave is unstable in the region outside the outer light surface,
while the Alfv\'{e}n wave is stable over the whole region in the case of
$\displaystyle \Omega_{\rm F} < \Omega_{\rm c} = \frac{2 M}{2M+r_{\rm H}}
|\Omega_{\rm H}|$. The instability of the Alfv\'{e}n wave may explain
the winding of the magnetic surface around the equatorial plane 
near the Kerr black hole, as shown in recent 
GRMHD simulations \citep{ripperda22}.

In this paper, we have used the stationary background solution 
in precisely the same way as
\citet{noda20} and we found that the Alfv\'{e}n wave is reflected
near the inner light surface but we did not identify 
the superradiance phenomenon.
This may be due to the initial perturbation conditions of the Alfv\'{e}n wave.

\begin{acknowledgments}

We are grateful to Mika Inda-Koide and Yasusada Nambu
for their helpful comments on this paper.
S.N. was supported by JSPS KAKENHI Grant No. 24K17053.
\end{acknowledgments}

\appendix

\section{The dispersion relation of an Alfv\'{e}n wave around a spinning black hole}
\label{appdx_disprel}

\begin{figure} 
\begin{center}
\includegraphics[width=14cm]{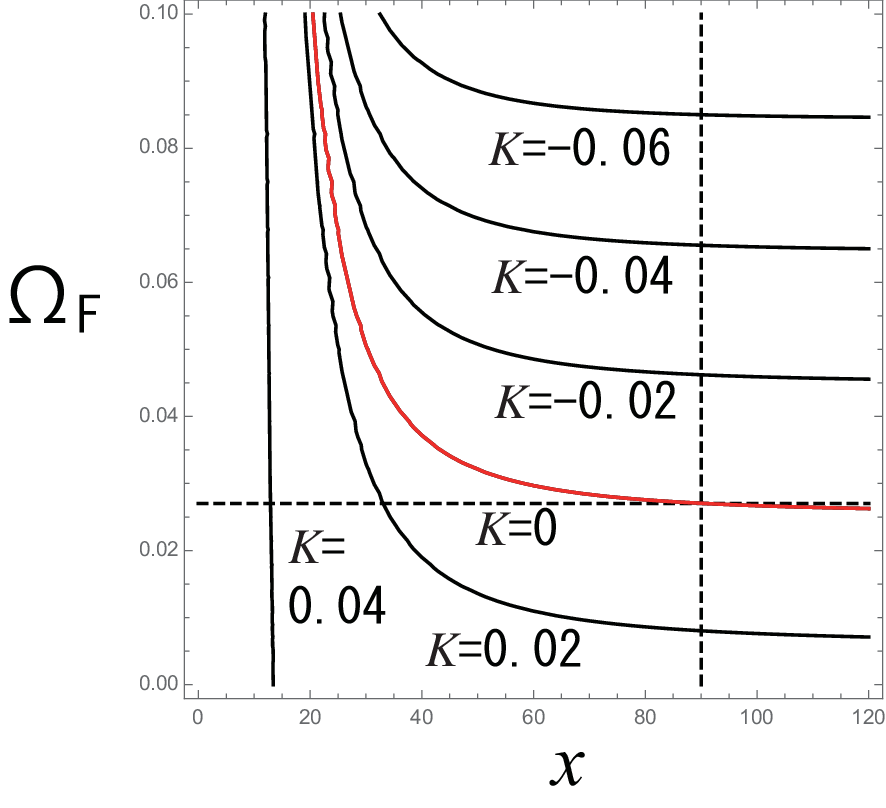}
\end{center}
\caption{
\textcolor{black}{
The propagation diagram of the Alfv\'{e}n wave with the plot of
$K$ in the case of the spin parameter $a=0.2$.
The solid red line shows the line of $K=0$. When $K$ is negative the propagation
of the Alfv\'{e}n wave is unstable, while the wave propagation is stable
when $K$ is positive.
The horizontal dashed line: $\Omega_{\rm F} = 0.027$.
The vertical dashed line: $x= 91$.
\label{fig:plot_K} 
}
}
\end{figure}

In this appendix, we discuss the strange behavior of an Alfv\'{e}n wave propagating along
a magnetic field line using the eikonal approximation.
The dispersion relation of the wave governed by Eq. (\ref{newtontoyeq}) is given by
\begin{equation}
\omega = \frac{V \pm \sqrt{V^2 + \lambda S}}{\lambda} k
= \frac{V \pm \sqrt{\Sigma K}/\sin \theta}{\lambda} k
\label{disrel4alfwv}
\end{equation}
in the small wavelength limit.
The dispersion relation shows that when $K$ is negative, the Alfv\'{e}n wave is unstable.
Hereafter, we consider a perturbation on the equatorial surface $\theta = 90^\circ$
($\Psi = 0$). At the infinity point ($X \longrightarrow \infty$),
we have 
\begin{equation}
K = \frac{\alpha^2 - R^2 (\Omega_{\rm F} - \Omega)^2 + I^2 X^2}{\Delta}
\longrightarrow \frac{X^2}{\Delta} \left [ \frac{2M}{r_{\rm H}} \Omega_{\rm H} -
\left ( \frac{2M}{r_{\rm H}-M} \right ) \Omega_{\rm F}  \right ]
\left [ \frac{2M}{r_{\rm H}} \Omega_{\rm H} -
\left ( \frac{2M}{r_{\rm H}+M} \right ) \Omega_{\rm F}  \right ].
\end{equation}
Then, in the case of 
$\displaystyle \Omega_{\rm c} \equiv \frac{2M}{2M +r_{\rm H}} | \Omega_{\rm H}|
< | \Omega_{\rm F} | < \frac{2M}{2M -r_{\rm H}} | \Omega_{\rm H}|$, 
$K$ becomes negative at a point far enough from the black hole 
($r \gg r_{\rm H}$); that is, the Alfv\'{e}n wave
in an outer region far enough from the hole is unstable.
\textcolor{black}{
In Fig. \ref{fig:plot_K}, we plot $K$ with respect to the position $x$ and $\Omega_F$
in the case of the spin parameter $a=0.2$. 
For the case of $\Omega_{\rm F} = 0.02$, the Alfv\'{e}n wave propagates
stably in the range $x \le 120$, while for $\Omega_{\rm F} = 0.027$
the Alfv\'{e}n wave becomes unstable at $x > 91.1$.}

\begin{figure} 
\begin{center}
\includegraphics[width=14cm]{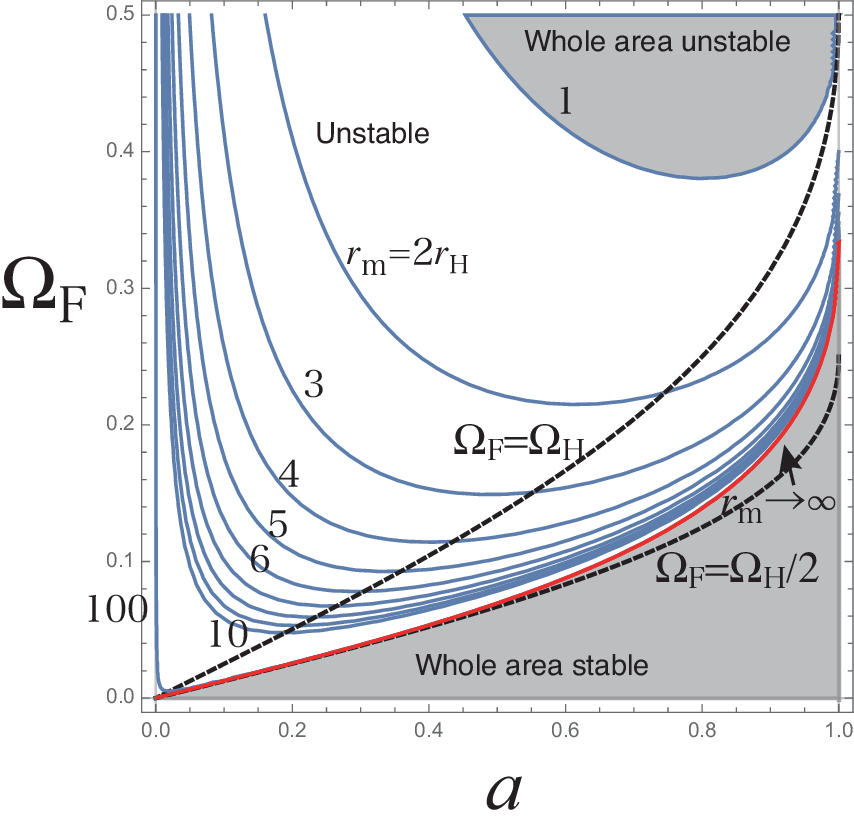}
\end{center}
\caption{The unstable region of Alfv\'{e}n wave propagation on the
equatorial plane for each $a$ and $\Omega_{\rm F}$.
The Alfv\'{e}n wave at $X > r_{\rm m}$ ($r > r_{\rm m}$) is unstable.
When $\Omega_{\rm F}$ is below the line of $r_{\rm m} \longrightarrow \infty$,
the Alfv\'{e}n wave is stable throughout the region from the horizon to infinity.
\label{dgm2alfinsta}}
\end{figure}

It is noted that an Alfv\'{e}n wave around a Schwarzschild black hole is
always stable because $\Omega_{\rm H} = 0$ and the unstable condition
of the wave can never be satisfied.
Figure \ref{dgm2alfinsta} shows the contour of $K=0$ at $X=r_{\rm m}$
for the case of $0 \le a \le 1$ and $0 \le |\Omega_{\rm F}| \le 0.5$.
$r_{\rm m}$ is the marginal radial coordinate of the instability
of the Alfv\'{e}n wave, where $r_{\rm m}$ is the solution for $K=0$.
The Alfv\'{e}n wave is unstable at $X > r_{\rm m}$ and stable at $X < r_{\rm m}$. 
The line of $\Omega_{\rm F} = \Omega_{\rm H}$ in Fig. \ref{dgm2alfinsta} shows that the 
Alfv\'{e}n wave is always unstable in the case of $\Omega_{\rm F} = \Omega_{\rm H}$,
which is surprising,
while the wave is stable in the case of $\Omega_{\rm F} = \Omega_{\rm H}/2$
because $\displaystyle \Omega_{\rm H}/2 < \frac{2M}{2M + r_{\rm H}} |\Omega_{\rm H}| \equiv \Omega_{\rm c}$.
For example, in the case of $a_* = a/M=0.2$ and $\Omega_{\rm F}=\Omega_{\rm H}$,
the Alfv\'{e}n wave is unstable in the region of $X > 9.5 r_{\rm H} = 18.8 M$.
However, the contours of $K=0$ confirm that the Alfv\'{e}n wave is stable when 
$\displaystyle \Omega_{\rm F} < \frac{2M}{2M +r_{\rm H}} | \Omega_{\rm H}| 
\equiv \Omega_{\rm c}$.
Hereafter, we mainly consider the propagating Alfv\'{e}n wave in the stable region
$X < r_{\rm m}$ or in the case of $\Omega_{\rm F} \ge \Omega_{\rm c}$.

The dispersion relation (\ref{disrel4alfwv}) shows that
the velocity of an Alfv\'{e}n wave governed by Eq. (\ref{newtontoyeq}) is
\begin{equation}
v_{\rm ph}^\pm = \frac{\omega}{k} = \frac{V \pm \sqrt{V^2 + \lambda S}}{\lambda}
= \frac{V \pm \sqrt{\Sigma K}/\sin \theta}{\lambda} 
\label{vphpm}
\end{equation}
when the wavelength is small enough.
Note also that Eq. (\ref{vphpm}) shows the propagation directions of the waves
with $v_+$ and $v_-$ are the same when $S$ is negative or equal to zero.
Between the outer and inner light surfaces, $S$ is positive.
However, outside of the region between the two light surfaces, $S$ is negative.
Thus, we conclude that the Alfv\'{e}n wave propagates only in one direction
outside the region between the inner and outer light surfaces.
This conclusion is also significant.

Here, we consider the head velocity of an outwardly propagating Alfv\'{e}n wave
at the outer light surface, 
which is given by
\begin{equation}
v_{\rm h, LS+} = \lim_{k \rightarrow \infty} \frac{\omega}{k} = v_{\rm ph}^+(r_{\rm LS}^+)
= \frac{V+|V|}{\lambda},
\end{equation}
because $S$ vanishes at the light surface.
The head velocity represents the upper limit of the propagation velocity of information.
When $0 < \Omega_{\rm F} < \Omega_{\rm H}$, the head velocity of the
Alfv\'{e}n wave at the outer light surface, $v_{\rm h, LS+}$ becomes
$\displaystyle \frac{2 V}{\lambda}$ because $V$ is positive.
Meanwhile, when $\Omega_{\rm F} < 0$ or $\Omega_{\rm F} > \Omega_{\rm H}$,
$v_{\rm h, LS+}$ vanishes because $V$ is negative.
This means the Alfv\'{e}n wave never passes 
through the outer light surface outward when $\Omega_{\rm F} < 0$ or 
$\Omega_{\rm F} > \Omega_{\rm H}$, while it does
when $0 < \Omega_{\rm F} < \Omega_{\rm H}$.
This results from the shape of the magnetic field line (Fig. \ref{pontie_zentai2}). 
The azimuthal component of the magnetic field in Boyer--Lindquist coordinates 
is given by 
$\displaystyle \bar{B}^\phi = ^\ast \hspace{-1pt} \bar{F}^{t\phi}=-\frac{I}{\Delta} = -\frac{2M}{\Delta r_{\rm H}} 
(\Omega_{\rm H} - \Omega_{\rm F})$, while the radial component of the magnetic field
is given by $\displaystyle \bar{B}^r = ^\ast \hspace{-1pt} \bar{F}^{tr}=\frac{1}{\Sigma} > 0$ 
at the equatorial plane $\displaystyle \theta = \frac{\pi}{2}$.
When $0 < \Omega_{\rm F} < \Omega_{\rm H}$, the magnetic field line bends
in the opposite direction to the rotation of the magnetic field line outward,
and the azimuthal component of the velocity of the Alfv\'{e}n wave
never exceeds the speed of light. Otherwise, the magnetic field line bends
in the opposite direction, and the wave velocity exceeds the speed of light if it passes
over the outer light surface.

\section{The energy-momentum tensor and 
force-free condition for an Alfv\'{e}n wave
\label{appenda}}
We show the detailed calculation with respect to the energy-momentum tensor and 
the force-free condition in the case
of the Alfv\'{e}n wave only, where we assume
\begin{equation}
\phi_1 = \bar{\phi}_1 + \delta \phi_1 = - \cos \theta + \psi, \verb!   !
\phi_2 = \bar{\phi}_2.
\end{equation}
First, we show non-zero components of the field tensor of the equilibrium, 
first and second-order perturbations as follows.
Non-vanishing components of the field tensor of the equilibrium are only
\begin{equation}
\bar{F}_{\rho \Psi} = - \bar{F}_{\Psi \rho} =-1, \verb!   !
\bar{F}^{\mu \Psi} = - \bar{F}^{\Psi \mu} = Y^{\mu\Psi\Psi} \verb!   ! (\mu \ne \Psi).
\end{equation}
The field tensor of the first-order perturbations are 
\begin{equation}
\delta F_{\mu\nu}  =  - \delta F_{\nu\mu} = 
(\delta^\rho_\nu - \Omega_{\rm F} \delta^T_\nu ) \partial_\mu \psi
- (\delta^\rho_\mu - \Omega_{\rm F} \delta^T_\mu) \partial_\nu \psi
\verb!   ! (\lambda \ne \rho) , \verb!   !
\delta F^{\mu \nu}  =   Y^{\mu\lambda\nu} \partial_\lambda \psi.
\end{equation}
All components of the field tensor of the second-order perturbations vanish.\\
Then, we calculate
\begin{eqnarray}
\bar{F}^{\mu\lambda} \delta F_{\nu \lambda} 
&=& \bar{F}^{\mu \Psi} \delta F_{\nu \Psi} = g^{\mu \rho} \delta F_{\nu \Psi} = 0, \\
\bar{F}^{\lambda\kappa} \delta F_{\lambda\kappa} 
&=& \bar{F}_{\lambda\kappa} \delta F^{\lambda\kappa}
= 2 \bar{F}_{\rho \Psi} \delta F^{\rho \Psi} = 2 W^{\rho\lambda \Psi \rho} \partial_\lambda \psi
= -2 g^{\rho\rho} g^{\lambda \Psi} \partial_\lambda \psi=0.
\end{eqnarray}
We find
\begin{eqnarray}
\delta F^{\mu\lambda} \bar{F}_{\nu\lambda}
&=& \delta F^{\mu \Psi} \bar{F}_{\nu \Psi} + \delta F^{\mu\rho} \bar{F}_{\nu\rho}
= \delta F^{\nu \Psi} \delta_{\nu\rho} - \delta^{\mu\rho} \delta_{\nu \Psi}
= W^{\mu\lambda \Psi\rho} \partial_{\lambda} \psi \delta_{\nu\rho}
- W^{\mu\lambda \rho\rho} \partial_{\lambda} \psi \delta_{\nu \Psi} \nonumber \\
&=& - g^{\mu\rho} g^{\Psi\Psi} \partial_\Psi \psi \delta_{\nu\rho}
- W^{\mu\lambda\rho\rho} \partial_\lambda \psi \delta_{\nu \Psi}
= -W^{\mu\lambda\rho\rho} \partial_\lambda \psi \delta_{\nu \Psi}=0.
\end{eqnarray}
Because when $\nu = \rho$, we have $\delta F_{\rho \Psi} = - \partial_\Psi \psi = 0$,
and otherwise, we have $\delta F_{\nu \Psi}=0$.
Then, we conclude that when $\mu$ and $\nu$ are not $\Psi$, we have
\begin{equation}
\delta T^\mu_\nu = 0.
\end{equation}

To check the force-free condition, we calculate the Lorentz force,
\begin{equation}
f_\mu^{\rm L} = J^\nu F_{\mu\nu}, \verb!   !
J^\mu = \nabla_\nu F^{\mu\nu} = \frac{1}{\sqrt{-g}} \partial_\nu (\sqrt{-g} F^{\mu\nu}).
\end{equation}
First, we can confirm the Lorentz force of the equilibrium vanishes because
the 4-current density of the equilibrium vanishes:
\begin{equation}
\bar{J}^\mu = \frac{1}{\sqrt{-g}} \partial_\nu (\sqrt{-g} \bar{F}^{\mu\nu})
= \frac{1}{\sqrt{-g}} \partial_X (\sqrt{-g} \bar{F}^{\mu X})=0,
\verb!   !
\bar{f}_\mu^{\rm L} = \bar{J}^\nu \bar{F}_{\mu\nu}=0,
\end{equation}

Second, we confirm the Lorentz force with respect to the first-order perturbation
of Alfv\'{e}n wave vanishes:
\begin{eqnarray}
\delta f_{i}^{\rm L} &=& \delta J^\nu \bar{F}_{i \nu} = 0, \\
\delta f_{\rho}^{\rm L} &=& \delta J^\nu \bar{F}_{\rho\nu} 
= \delta J^\Psi \bar{F}_{\rho \Psi} = \delta J^\Psi
= \frac{1}{\sqrt{-g}} \partial_\nu (\sqrt{-g} \bar{F}^{\Psi \nu})
= \frac{1}{\sqrt{-g}} \partial_\nu (\sqrt{-g} Y^{\Psi \lambda\nu} \partial_\lambda \psi)
= 0, \\
\delta f_{\Psi}^{\rm L} &=& \delta J^\nu \bar{F}_{\Psi \nu} 
= \delta J^\rho \bar{F}_{\Psi \rho} = - \delta J^\rho
= - \frac{1}{\sqrt{-g}} \partial_\nu (\sqrt{-g} \delta F^{\rho \nu}) \nonumber \\
& = & -\frac{1}{\sqrt{-g}} \partial_\nu (\sqrt{-g} Y^{\rho\lambda\nu} \partial_\lambda \psi)
= \frac{1}{\sqrt{-g}} \partial_\nu (\sqrt{-g} Z^{\nu\lambda} \partial_\lambda \psi)
= 0,
\label{forcefreecheck}
\end{eqnarray}
where $i$ is $T$ or $X$.
In the last equation of Eq. (\ref{forcefreecheck}), we use Eq.
(\ref{nodaeq16d}).
Then, we confirmed the first order of the Lorentz force vanishes.

It is noted that we have $\delta J^\rho = \delta J^\Psi = 0$.
Using
\begin{equation}
\delta J^\mu = \frac{1}{\sqrt{-g}} \partial_\nu (\sqrt{-g} \delta F^{\mu\nu})
=  \frac{1}{\sqrt{-g}} \partial_\nu (\sqrt{-g} Y^{\mu\lambda\nu} \partial_\lambda \psi),
\end{equation}
we have 
\begin{eqnarray}
\delta J^T & =& \frac{1}{\sqrt{-g}} \sum_{\lambda = T, X} 
\partial_\nu (\sqrt{-g} Y^{T \lambda\nu} \partial_\lambda \psi)
= \frac{1}{\sqrt{-g}} \sum_{\lambda = T, X} 
\partial_X (\sqrt{-g} Y^{T \lambda X} \partial_\lambda \psi) ,\\
\delta J^X & =& \frac{1}{\sqrt{-g}} \sum_{\lambda = T, X} 
\partial_\nu (\sqrt{-g} Y^{X \lambda\nu} \partial_\lambda \psi)
= - \frac{1}{\sqrt{-g}} \sum_{\lambda = T, X} 
\partial_X (\sqrt{-g} Y^{T \lambda X} \partial_\lambda \psi), \\
\delta J^\rho &=& \delta J^\Psi = 0,
\end{eqnarray}
where $\displaystyle Y^{TTX} = \frac{I}{\Sigma}$ and 
$Y^{TXX} = W = \Omega - \Omega_{\rm F}$.

We also have
\begin{equation}
\delta^2 J^\mu = \frac{1}{\sqrt{-g}} \partial_\nu (\sqrt{-g} \delta^2 F^{\mu\nu})=0 .
\end{equation}
We find three components of the  Lorentz force vanish as
\begin{eqnarray}
\delta^2 f_{T}^{\rm L} &=& \delta J^\mu \delta F_{T \mu} 
= \frac{1}{\sqrt{-g}} \partial_\nu (\sqrt{-g} Y^{\mu\lambda\nu}) 
\Omega_{\rm F} \partial_\mu \psi ,\\
\delta^2 f_{X}^{\rm L} &=& \delta J^\mu \delta F_{X \mu} =  0,\\
\delta^2 f_{\Psi}^{\rm L} &=& \delta J^\mu \delta F_{\Psi \mu} =  0 ,
\end{eqnarray}
because of $\delta J^\rho = 0$.
The $\rho$-component of the Lorentz force is calculated as,
\begin{eqnarray}
\delta^2 f_{\rho}^{\rm L} &=& \delta J^\mu \delta F_{\rho\mu} 
= \frac{1}{\sqrt{-g}} \partial_\nu (\sqrt{-g} Y^{\mu\lambda\nu} \partial_\lambda \psi) 
\partial_\mu \psi \nonumber \\
&=& \frac{1}{\sqrt{-g}} \sum_{\lambda = T, X} 
[ - \partial_X (\sqrt{-g} Y^{T \lambda X} \partial_\lambda \psi) \partial_T \psi
+ \partial_T (\sqrt{-g} Y^{T \lambda X} \partial_\lambda \psi) \partial_X \psi ] \nonumber \\
&=& \frac{1}{\sqrt{-g}}
\left [ - \partial_X ( \Theta \partial_T \psi )
 + \partial_T ( \Theta \partial_X \psi ) \right ] ,
\label{appdx_djrhol}
\end{eqnarray}
where $\displaystyle \Theta = \sqrt{-g} Y^{T\alpha X} \partial_\alpha \psi
= - \frac{I X}{\alpha^2} \partial_T \psi - R^2 (\Omega_{\rm F} - \Omega) \partial_X \psi$.
Here, we have the relation,
\begin{equation}
\delta^2 f_{\rm L}^T = \Omega_{\rm F} \delta^2 f_{\rm L}^\rho.
\end{equation}
To vanish $\delta^2 f_\rho^{\rm L}$ over the entire radial range ($X > r_{\rm H}$), $\Theta$
should be zero. When $\Theta$ is zero, $I$, $\Omega_{\rm F}$, and $\Omega$ are zero.
Then, only when $I$, $\Omega_{\rm F}$, and $a_\ast$ are all zero, $\delta^2 f_\rho^{\rm L}$
vanishes. Then, except for the case of $I$, $\Omega_{\rm F}$, $a_\ast$, the Alfv\'{e}n 
wave induces the fast wave.

\section{The energy density and energy flux density
corrected by $\chi$
\label{appendb}}

We show the details of the calculation of the energy density and energy flux
of Alfv\'{e}n wave and the induced fast wave described by $\chi$.
The Alfv\'{e}n wave and the fast wave are given by
the following perturbation, respectively,
\begin{eqnarray}
\phi_1 &=& \bar{\phi}_1 + \delta \phi_1 = \Psi + \psi(T, X), \\
\phi_2 &=& \bar{\phi}_2 + \delta^2 \phi_2 = \rho - \Omega_{\rm F} T + \chi(T, X) .
\end{eqnarray}
To derive time evolution equation of $\chi$, we use Eq. (\ref{nodaeq04w}).
When we take $i=2$, we find the trivial equation:
\begin{equation}
\partial_\lambda (\sqrt{-g} W^{\lambda\alpha\mu\beta} \partial_\alpha \bar{\phi}_1
\partial_\beta \chi ) \partial_\mu \bar{\phi}_2
+ \partial_\lambda (\sqrt{-g} W^{\lambda\alpha\mu\beta} \partial_\alpha \bar{\phi}_1
\partial_\beta \bar{\phi}_2 ) \partial_\mu \chi
= - \partial_\lambda (\sqrt{-g} W^{\lambda z \rho\beta} \partial_\beta \chi)
- \partial_\lambda (\sqrt{-g} W^{\lambda z \mu\rho} \partial_\mu \chi) = 0.
\end{equation}
When we take $i=1$, we obtain the equation of $\chi$:
\begin{eqnarray}
&& \partial_\lambda (\sqrt{-g} W^{\lambda\alpha\mu\beta} \partial_\alpha \psi
\partial_\beta \bar{\phi}_2 ) \partial_\mu \psi
+ \partial_\lambda (\sqrt{-g} W^{\lambda\alpha\mu\beta} \partial_\alpha \bar{\phi}_1
\partial_\beta \chi) \partial_\mu \bar{\phi}_1     \nonumber \\
&=& \partial_\lambda (\sqrt{-g} Y^{\lambda\alpha\mu} \partial_\alpha \psi) 
\partial_\mu \psi
+ \partial_\lambda (\sqrt{-g} W^{\lambda \Psi\Psi \beta} \partial_\beta \chi)  \nonumber \\
&=& \partial_\lambda (-\sqrt{-g} g^{\lambda\beta} \partial_\beta \psi) 
+ \partial_\lambda (\sqrt{-g} W^{\lambda\alpha\mu\rho} \partial_\alpha \psi) 
\partial_\mu \psi = 0.
\end{eqnarray}
Then, we obtain the time evolution equation of $\chi$,
\begin{equation}
\Box (g^{\Psi\Psi} \chi) = \nabla_\mu \nabla^\mu (g^{\Psi\Psi} \chi) 
= \frac{1}{\sqrt{-g}} \partial_\lambda
(\sqrt{-g} g^{\Psi\Psi} g^{\mu\beta} \partial_\beta \chi)
= \frac{1}{\sqrt{-g}} \partial_\lambda (\sqrt{-g} Y^{\lambda\alpha\mu}
\partial_\alpha \psi \partial_\mu \psi) \equiv s,
\label{eqchi4dlb_ap}
\end{equation}
where $\Box \equiv \nabla_\mu \nabla^\mu$ is the d'Alembertian.
When we consider $\chi$, we found the force-free condition recovers as follows.
When we take $\nu \ne \rho$, we have
\begin{equation}
\delta^2 f_\nu = \delta J^\mu \delta F_{\nu \mu} + \delta^2 J^\mu \bar{F}_{\nu \mu}
= \delta J^\rho \delta F_{\nu \rho} + \delta^2 J^\rho \bar{F}_{\nu \rho} = 0. 
\end{equation}
Otherwise, we have
\begin{align}
\delta^2 f_\rho &= \delta J^\mu \delta F_{\rho \mu} + \delta^2 J^\mu \bar{F}_{\rho \mu}
= - \frac{1}{\sqrt{-g}} \partial_\nu (\sqrt{-g} Y^{\mu\lambda\nu} \partial_\lambda \psi)
\partial_\mu \psi + \delta^2 J^z  \nonumber \\
& = - \frac{1}{\sqrt{-g}} \partial_\nu (\sqrt{-g} g^{\Psi\Psi} g^{\nu\beta} \partial_\beta \chi)
+ \frac{1}{\sqrt{-g}} \partial_\nu (\sqrt{-g} Y^{\nu\lambda\mu} \partial_\lambda \psi
\partial_\mu \psi) = 0.
\end{align}
Eventually, introducing $\chi$, we recovered and confirmed the force-free condition
up to the second order perturbation.

We show the detail derivation of the equilibrium, first order, and second order
of the values with respect to the conservation law.
With respect to the energy density, we have
\begin{eqnarray}
\bar{S}^T &=& -\frac{1}{2} \bar{F}^{Ti} \bar{F}_{Ti} 
+ \frac{1}{4} \bar{F}^{ij} \bar{F}_{ij} 
= \frac{1}{2} (Y^{\Psi\Psi\rho}+\Omega_{\rm F} Y^{\Psi\Psi T})
= \frac{1}{2 \Delta \Sigma} \left [ \alpha^2 - R^2 (\Omega^2 - \Omega_{\rm F}^2)
+ \frac{1}{\Sigma} I^2 X^2 \sin^2 \theta \right ]  \\
\delta S^T &=&  -\frac{1}{2} \delta F^{Ti} \bar{F}_{Ti} 
- \frac{1}{2} \bar{F}^{Ti} \delta F_{Ti}
+ \frac{1}{4} \delta F^{ij} \bar{F}_{ij}+ \frac{1}{4} \bar{F}^{ij} \delta F_{ij} 
=\left (Y^{\Psi\Psi\rho}+\frac{1}{2} \Omega_{\rm F} Y^{\Psi\Psi T} \right )\partial_\Psi \psi
=  0, \\
\delta^2 S^T &=&  -\frac{1}{2} \delta F^{Ti} \delta F_{Ti} 
+ \frac{1}{4} \delta F^{ij} \delta F_{ij} 
-\frac{1}{2} \bar{F}^{iT} \delta^2 F_{Ti} + \frac{1}{4} \bar{F}^{ij} \delta^2 F_{ij}
+ \frac{1}{4} \delta^2 F^{ij} \bar{F}_{ij}  \nonumber \\
&=& - \frac{1}{2} [ Y^{T \lambda\rho} \partial_T \psi + (Y^{\rho\lambda i} + \Omega_{\rm F}
Y^{T \lambda i}) \partial_i \psi ] \partial_\lambda \psi
+ Y^{\Psi\Psi j} \partial_j \chi + \Omega_{\rm F} W^{\Psi\Psi\kappa T} \partial_\kappa \chi
\nonumber \\
&=& \frac{1}{\Sigma} \left [
\frac{1}{2} \lambda (\partial_T \psi)^2
+ \frac{1}{2} \{ \alpha^2 + R^2 (\Omega_{\rm F}^2 - \Omega^2) \}
(\partial_X \psi)^2  \right ]
+ \frac{1}{\alpha^2} \Omega_{\rm F} I \partial_X \psi \partial_T \psi
+ \frac{1}{\Sigma} (I \partial_X \chi - \frac{1}{\alpha^2} 
\Omega_{\rm F} \partial_T \chi)
\end{eqnarray}
With respect to the energy flux, we have 
\begin{eqnarray}
\bar{S}^X &=& \bar{F}^{Xi} \bar{F}_{iT} = 
\Omega_{\rm F} Y^{\Psi\Psi X}
= \frac{X^2}{\Sigma^2} I \Omega_{\rm F} \sin^2 \theta,  \\
\delta S^X &=&  \delta F^{Xi} \bar{F}_{iT} + \bar{F}^{Xi} \delta F_{iT}
= \Omega_{\rm F}^2 Y^{\Psi\Psi X} \partial_\Psi \psi  = 0, \\
\delta^2 S^X &=&  \delta F^{Xi} \delta F_{iT} 
+ \bar{F}^{Xi} \delta^2 F_{iT}
=  \delta F^{X\rho} \delta F_{\rho T}
+\bar{F}^{X\Psi} \delta^2 F_{\Psi T} \nonumber \\
&=& - Y^{X \lambda \rho} \partial_\lambda \psi \partial_T \psi  
- Y^{\Psi\Psi X} \partial_T \chi + 
\Omega_{\rm F} W^{\Psi\Psi X \kappa} \partial_\kappa \chi \nonumber \\
&=& - \frac{1}{\Sigma} \partial_T \psi
\left [ \frac{2 M a I X}{\Delta} \partial_T \psi
+\{ \alpha^2 + R^2 \Omega (\Omega_{\rm F} - \Omega) \partial_X \psi \right ]
+ \frac{1}{\Sigma} \left [ - I \partial_T \chi 
+ \frac{\Delta}{X^2} \Omega_{\rm F} \partial_X \chi \right ].
\end{eqnarray}

With respect to the angular momentum, we introduced $\chi$
to recover the force-free condition of the Alfv\'{e}n wave
up to second order of the perturbation so that we become able to consider 
the conservation law of angular momentum up to second order.
The axial Killing vector $\xi^\mu_{(\rho)} = (0, 0, 1, 0)$ yields
the 4-angular momentum flux density $M^\mu = \xi^\nu_{(\rho)} T^\mu_\nu$, and
we obtain the angular momentum conservation law in the corotating natural coordinates,
\begin{equation}
\nabla_\mu M^\mu = \frac{1}{\sqrt{-g}} \partial_\mu (\sqrt{-g} M^\mu)
= \frac{\partial M^0}{\partial T} + \frac{1}{\sqrt{-g}} \frac{\partial}{\partial X}
(\sqrt{-g} M^X) = \xi^\mu_{(\rho)} f_\mu^{\rm L} = f_\rho^{\rm L} = 0 .
\end{equation}
The 4-angular momentum density $M^\mu$
in the corotating natural coordinates calculated as
\begin{equation}
M^\mu = \xi^\nu_{(\rho)} T^{\mu}_{\nu} = T^\mu_\rho = F^{\mu \nu} F_{\rho\nu}.
\end{equation}
Then, we have the 4-angular momentum density of the equilibrium, first and
second-order perturbations of the linear Alfv\'{e}n wave,
\begin{eqnarray}
\bar{M}^\mu &=& \bar{T}^\mu_\rho = \bar{F}^{\mu\lambda} \bar{F}_{\rho\lambda} 
= Y^{\Psi\Psi\mu}, \\
\delta M^\mu &=& 
\delta F^{\mu\nu} \bar{F}_{\rho\nu} + \bar{F}^{\mu\nu} \delta F_{\rho\nu}
= \delta F^{\mu \Psi} \bar{F}_{\rho \Psi} + \bar{F}^{\mu \Psi} \delta F_{\rho \Psi}
= - \delta F^{\mu \Psi} - Y^{\Psi\Psi\mu} \partial_\Psi \psi = 0, \\
\delta^2 M^\mu &=& \delta^2 F^{\mu\nu} \bar{F}_{\rho\nu} 
+ \delta F^{\mu\nu} \delta F_{\rho\nu} + \bar{F}_{\mu\nu} \delta^2 F^{\rho\nu}
= \delta^2 F^{\mu \Psi} \bar{F}_{\rho \Psi} 
+ \bar{F}^{\mu \Psi} \delta^2 F_{\rho \Psi} + \delta F^{\mu \nu} \delta F_{\rho\nu} \nonumber \\
& = & - \delta^2 F^{\mu\Psi} - Y^{\Psi\Psi\mu} \delta^2 F_{\rho\Psi}
+ \delta F^{\mu\nu} \delta F_{\rho\nu} \nonumber \\
&=& Y^{\mu\lambda\nu} \partial_\lambda \psi 
(- \Omega_{\rm F} \delta^T_\nu \partial_\rho \psi - \partial_\nu \psi ) 
+ W^{\Psi\Psi\mu\kappa} \partial_\kappa \chi
+ Y^{\Psi\Psi\mu} \partial_\rho\chi
= -Y^{\mu\lambda\nu} \partial_\lambda \psi \partial_\nu \psi
+ W^{\Psi\Psi\mu\kappa} \partial_\kappa \chi ,
\label{angmomexp}
\end{eqnarray}
where we assume $\mu \ne \Psi$.
Then, we confirm the conservation of the angular momentum for the second 
order of the perturbation as
\begin{equation}
\nabla_\mu \delta^2 M^\mu = \frac{1}{\sqrt{-g}} \partial_\lambda (\sqrt{-g} \delta^2 M^\lambda)
= \frac{1}{\sqrt{-g}} \partial_\lambda [\sqrt{-g} 
(Y^{\nu\lambda\mu} \partial_\lambda \psi \partial_\nu \psi
+W^{\Psi\Psi\mu\kappa} \partial_\kappa \chi )] = 0,
\end{equation}
where we used Eq. (\ref{angmomexp}).
We have the energy density and energy flux density of the equilibrium, first and
second-order perturbations of the linear Alfv\'{e}n wave distinctively,
\begin{eqnarray}
\bar{M}^T &=& Y^{\Psi\Psi T} , \verb!   !
\bar{M}^X =  Y^{\Psi\Psi X} , \\
\delta M^T &=&  0, \verb!   !
\delta M^X =  0, \\
\delta^2 M^T &=& 
- Y^{T\lambda\nu} \partial_\lambda \psi \partial_\nu \psi 
+ W^{\Psi\Psi T\kappa} \partial_\kappa \chi
= - Y^{T\lambda X} \partial_\lambda \psi \partial_X \psi 
+ g^{\Psi\Psi} g^{T\kappa} \partial_\kappa \chi ,\\
\delta^2 M^X &=& 
- Y^{X\lambda\nu} \partial_\lambda \psi \partial_\nu \psi 
+ W^{\Psi\Psi X\kappa} \partial_\kappa \chi
= Y^{T\lambda X} \partial_\lambda \psi \partial_T \psi 
+ g^{\Psi\Psi} g^{X \kappa} \partial_\kappa \chi .
\end{eqnarray}

\section{The energy transport with waves \label{appendetbtz}}
We recovered the force-free condition with
the additional variable $\chi$ so that we have the
energy and momentum conservation,
$\nabla_{{\nu}} T^{{\mu}{\nu}} = 0$
up to the second order of the perturbations.
When we use the time-like Killing vector $\xi_{(T)}^{{\mu}} = (1, 0, 0, 0)$,
we have the energy conservation law,
\begin{equation}
\nabla_{{\nu}} {S}^{{\nu}}
= \frac{1}{\sqrt{-g}} \partial_{{\nu}} ( \sqrt{-g} {S}^{{\nu}})
= \frac{1}{\sqrt{-g}} \partial_T ( \sqrt{-g} {S}^T)
+ \frac{1}{\sqrt{-g}} \partial_{{i}} ( \sqrt{-g} {S}^{{i}}) = 0 ,
\end{equation}
where 
${S}^{{\nu}} = - \xi_{{\mu}}^{(T)} T^{{\mu}{\nu}}$
is the 4-energy flux density.
In the case of axisymmetry and translation symmetry with respect to the $\Psi$-direction,
we have
\begin{equation}
\frac{\partial}{\partial T} ( \sqrt{-g} {S}^T)
+ \frac{\partial}{\partial X} ( \sqrt{-g} {S}^X)
+ \frac{\partial}{\partial \Psi} (\sqrt{-g} \delta^2 S^\Psi) = 0 .
\label{eq2ene}
\end{equation}
Here, we have $\delta^2 S^\Psi = \delta^{1+1} S^\Psi = - \delta F^{\Psi j} \delta F_{Tj}
= - Y^{\Psi\lambda j} \partial_\lambda \psi (\delta^\rho_j \partial_T \psi
+ \Omega_{\rm F} \partial_j \psi) = - Y^{\Psi\Psi j} \partial_\Psi \psi
(\delta^\rho_j \partial_T \psi + \Omega_{\rm F} \partial_j \psi)
= - (Y^{\Psi\Psi\rho} \partial_T \psi + \Omega_{\rm F} Y^{\Psi\Psi X}
\partial_X \psi) \partial_\Psi \psi$, and Eq. (\ref{eq2ene}) and $\partial^2_\Psi \psi = - \psi$
yield the conservation expression,
\begin{equation}
\frac{\partial}{\partial T} (\sqrt{-g} \delta^2 e^\infty)
+ \frac{\partial}{\partial X} (\sqrt{-g} \delta^2 S_{\rm P}) = 0,
\end{equation}
where $\displaystyle \delta^2 e^\infty = \delta^2 S^T + \frac{1}{2} Y^{\Psi\Psi\rho} \psi^2$ 
and $\displaystyle \delta^2 S_{\rm P} = \delta^2 S^X + \frac{1}{2 \sqrt{-g}} \Omega_{\rm F} I \psi^2$.
It reads,
\begin{align}
& \int_{T_1}^{T_2} dT \int_{X_1}^{X_2 } \frac{\partial}{\partial T} ( \sqrt{-g} \delta^2 e^\infty)
+ \int_{T_1}^{T_2} dT \int_{X_1}^{X_2 } \frac{\partial}{\partial X} ( \sqrt{-g} \delta^2 S_{\rm P})
\nonumber \\
& =   \int_{X_1}^{X_2 } \sqrt{-g} \delta^2 e^\infty(X,T_2)
-  \int_{X_1}^{X_2 } \sqrt{-g} \delta^2 e^\infty(X,T_1)
+ \int_{T_1}^{T_2} dT \sqrt{-g} \delta^2 S_{\rm P}(X_2,T) 
- \int_{T_1}^{T_2} dT \sqrt{-g} \delta^2 S_{\rm P}(X_1,T) = 0.
\end{align}
When we define the total energy between $X=X_1$ and $X=X_2$ and
the energy flux at $X=X_b$ ($b=1, 2$) by
\begin{eqnarray}
\delta^2 E(T) & = & \int_{X_1}^{X_2} \sqrt{-g} \delta^2 e^\infty(X, T) dX, 
\label{eqofet} \\
\delta^2 F_b (T) & = & \int_{T_1}^{T} \sqrt{-g} \delta^2 S_{\rm P}(X_b, T') dT',
\label{eqoffb}
\end{eqnarray}
respectively, we obtain the conservation quantity as
\begin{equation}
\delta^2 E(T) - \delta^2 F_1 (T) + \delta^2 F_2 (T) = \delta^2 E(0).
\end{equation}
Note that the energy of the Alfv\'{e}n wave is conserved
in the corotating magnetic natural frame $(T', X', \Psi', \rho')$,
\begin{equation}
\delta^{1+1} E'(T) - \delta^{1+1} F'_{1} (T) 
+ \delta^{1+1} F'_{2} (T) = \delta^{1+1} E'(0),
\end{equation}
where $\displaystyle \delta^{1+1} E'(T) 
= \int_{X_1}^{X_2} \sqrt{-g} \delta^{1+1} e^{\infty\prime} (X, T) dX$ and
$\displaystyle \delta^{1+1} F'_b (T) 
= \int_{0}^{T} \sqrt{-g} \delta^{1+1} S_{\rm P}^{\prime} (X_b, T') dT'$
($b=1, 2$).

\section{The numerical method used for the 1-D wave equation \label{append_numeth}}
\label{appendnum}

Equations (\ref{newtontoyeq}) with the additional term $-\kappa(x) \psi$ 
in its right-hand side and Eq. (\ref{chieq}) are written 
by multi-dimensional time-development equations as,
\begin{eqnarray}
\frac{\partial \psi}{\partial T} &=& 
- \frac{1}{\lambda(x)} \frac{\partial}{\partial x} (\xi + 2 V \psi)
- \frac{K}{\lambda} \zeta+ \frac{1}{\lambda} \frac{\partial V}{\partial X} \psi, 
\nonumber \\
\frac{\partial \xi}{\partial T} &=& - S \frac{\partial \psi}{\partial X} , \label{multdimeq} \\
\frac{\partial \zeta}{\partial T} &=& \psi, \nonumber \\
\frac{\partial \chi}{\partial T} &=& - \alpha^2 \frac{\partial \upsilon}{\partial X}
+ \sigma, \nonumber \\
\frac{\partial \upsilon}{\partial T} &=& - \frac{\Delta}{\Sigma} 
\frac{\partial \chi}{\partial X}, \nonumber \\
\frac{\partial \sigma}{\partial T} &=& g(X), \nonumber
\end{eqnarray}
where $\xi(X,T)$, $\zeta(X,T)$, $\upsilon(X,T)$, and $\sigma(X,T)$ are new variables and
$\displaystyle g(x) = \frac{- \alpha^2 \sqrt{-g}}{g^{TT}} s$.

We use the two-step Lax-Wendroff scheme for the multi-dimension
time-development equation
\begin{equation}
\frac{\partial \VEC{u}}{\partial t} =
- \VEC{h} \odot \frac{\partial \VEC{w}}{\partial X} + \VEC{f},
\label{multieqn}
\end{equation}
where $\VEC{u}$ is the array of the conserved quantity density, 
$\VEC{w}$ is the array of the flux density of the conserved quantity, 
$\VEC{f}$ is the array of the source density of the conserved variable:
\begin{eqnarray}
\VEC{u}_j^{\overline{n+1}} &=& \VEC{u}_j^n - \frac{\Delta t}{\Delta x} \VEC{h}_j^n \odot
(\VEC{w}_{j+1}^n - \VEC{w}_j^n) + \Delta t \VEC{f}_j^n ,\\
\VEC{u}_j^{n+1} &=& \frac{1}{2} \left [ \VEC{u}_j^n  \VEC{u}_j^{\overline{n+1}}
- \frac{\Delta t}{\Delta x} \VEC{h}_j^n \odot (\VEC{w}_j^{\overline{n+1}} - \VEC{w}_{j-1}^{\overline{n+1}}) 
+ \Delta t \VEC{f}_j^{\overline{n+1}} \right ] .
\end{eqnarray}
Here, we used $\odot$ to express the product of two vectors 
$\VEC{a}=(a_1, a_2, \cdots)^{\rm T}$ and
$\VEC{b}=(b_1, b_2, \cdots)^{\rm T}$,
\begin{equation}
\VEC{a} \odot \VEC{b} \equiv 
\left ( \begin{array}{c} a_1 b_1 \\ a_2 b_2 \\ \vdots \end{array} \right ).
\end{equation}

Equation (\ref{multdimeq}) are given by
\begin{equation}
\VEC{u} = \left ( \begin{array}{c} \psi \\ \xi \\ \zeta \\ 
\chi \\ \upsilon \\ \sigma \end{array} \right ),
\VEC{h} = \left ( \begin{array}{c} \frac{1}{\lambda(x)} \\ S \\ 1 \\
\alpha^2 \\ \frac{\Delta}{\Sigma} \\ 1 \end{array} \right ),
\VEC{w} = \left ( \begin{array}{c} \xi +2 V \psi \\ \psi \\ 0 \\
\upsilon \\ \chi \\ 0 \end{array} \right ),
\VEC{f} = \left ( \begin{array}{c} - \frac{1}{\lambda} \left ( \zeta- \frac{\partial V}{\partial X} \psi \right ) \\ 0 \\ \psi \\
\sigma \\ 0 \\ g  \end{array} \right ).
\end{equation}
To perform the precise calculation near the horizon, we use the tortoise coordinates, $x$
defined by $\displaystyle \frac{dX}{dx} = 1 - \frac{r_{\rm H}}{X}$.
In the calculations in this paper, we use the tortoise coordinate, $x = X - 2M +
r_{\rm H} \log(X-r_{\rm H})/(2M-r_{\rm H})$, where the $x=0$ corresponds to the
static limit ($X=2M$). The multidimensional equation (\ref{multieqn}) is replaced by 
$\displaystyle \frac{\partial \VEC{u}}{\partial T} = - \frac{dx}{dX} \VEC{h}
\odot \frac{\partial}{\partial X} \VEC{w} + \VEC{f}$, and then $\VEC{h}$
of Eq. (\ref{multieqn}) is replaced by $\displaystyle \frac{dx}{dX} \VEC{h}$.
\end{document}